\documentstyle[amssymb,aps,multicol,epsf]{revtex}

\begin{document}
\draft

\title{
Scaling properties of scale-free evolving networks: Continuous approach
}

\author{
S.N. Dorogovtsev$^{1, 2, \ast
}$ and J.F.F. Mendes$^{1,\dagger}$
}

\address{
$^{1}$ Departamento de F\'\i sica and Centro de F\'\i sica do Porto, Faculdade 
de Ci\^encias, 
Universidade do Porto\\
Rua do Campo Alegre 687, 4169-007 Porto, Portugal\\
$^{2}$ A.F. Ioffe Physico-Technical Institute, 194021 St. Petersburg, Russia 
}

\maketitle

\begin{abstract}
Scaling behavior of scale-free evolving networks arising in communications, citations, collaborations, etc. areas 
is studied. We derive universal scaling relations describing properties of such networks and indicate limits of their validity.
We show that main properties of scale-free evolving networks may be described in frames of a simple continuous approach. The simplest models of networks, which growth is determined by a mechanism of preferential linking, are used. We consider different forms of this preference and demonstrate that the range of types of preference linking producing scale-free networks is wide. We obtain also scaling 
relations for networks with nonlinear, accelerating growth and describe temporal evolution of arising distributions. Size-effects -- cut-offs of these distributions -- implement restrictions for observation of power-law dependences. The main characteristic of interest is so-called degree distribution, i.e., distribution of a number of connections of nodes. A scaling form of the distribution of links between pairs of individual nodes for the growing network of citations is also studied. We describe effects that produce difference of nodes. ``Aging'' of nodes changes exponents of distributions. Appearence of a single ``strong'' node changes dramatically the degree distribution of a network. 
If its strength exceeds some threshold value, the strong node captures a finite part of all links 
of a network. We show that permanent random damage of a growing scale-free network -- permanent deleting of some links -- change radically values of the scaling exponents. 
We describe the arising rich phase diagram. Results of other types of permanent damage are described.      
\end{abstract}


\begin{multicols}{2}

\narrowtext

%%%%%%%%%%%%%%%%%%%%%%%%%%%%%%%%%%%%%%%%%%%%%%%%%%%%%%%%%%%%%%%%%%%%%%%

%%%%%%%%%%
%%%%%%%%%%
%%%%%%%%%%
%%%%%%%%%%

\section{Introduction}\label{s-introduction} 

One of the most impressive recent discoveries in the field of the network evolution is an observation that a 
number of large growing networks are scale-free, that is distributions of numbers of connections of their nodes 
are of a power-law form \cite{ajb99,ha99,krrt99,bkm00,ba99,fff99}. To the class of scale-free networks 
belongs huge communications networks (World Wide Web and Internet \cite{b1,hppj98}), networks of citations in literature \cite{ls98,r98}, collaboration networks \cite{asbs00,mn00}, some biological networks (nets of metabolic reactions \cite{jta00}), etc. 
In fact, an incredible progress of science and information technology itself has produced networks, 
large enough (World Wide Web has about $10^9$ nodes) to get reliable data. 

Nevertheless, the experimental data are certainly not excellent.  
Indeed, the task of observation of such power laws is not easy. Arising distributions are very sensible to size-effects. 
Unfortunately, few really large growing networks are known yet, many of obtained data for not so large nets are very preliminary. It is a great temptation to describe any decreasing experimental curve in a log-log plot by a linear fit. Nearly all the observations are for the only quantity -- degree distribution of nodes. (Following definitions of mathematicians and computer scientists, we call here a total number of connections of a node its {\em degree} \cite{note1}. If the links are connected, a number of incoming links of a node is called {\em in-degree} of it, and a number of outgoing links -- {\em out-degree}.)  
Information about others characteristics is much poorer. 

Therefore, only a few scale-free growing networks have been observed. 
Why are these observations so important? Why the interest in these networks is so great? 
(In one of the last issues of Physical Review Letters, three papers, one after the other, were devoted to scale-free networks \cite{ceah00,krl00,dms00}.) Of course, the reason is not only power-law dependences of distributions itself but a variety of intriguing properties of scale-free networks which explain their existence in Nature. Degree distribution is a simple but very important characteristic of a network. In particular, it was shown that networks with power-law degree distributions are extremely resilient against random breakdowns if an exponent $\gamma$ of their degree distribution 
does not exceed $3$. This property is vitally important for communications and biological networks. One has also point out the following circumstance.  
A power-law dependence of degree distributions indicates scale-invariance of characteristics of 
networks and their hierarchically organized structure. Just investigation of scaling properties of scale-free networks and connections between them is a topic of the present paper. 

We study the formulated problem applying two different approaches in arbitrary order. 
First, we obtain universal scaling relations using general considerations. 
Second, we demonstrate features of scaling behavior of networks using minimal models providing the effect being studied. At the moment, the only known mechanism providing scale-free networks is preferential linking -- new links are preferentially attached to nodes with large number of connections (large degree) \cite{ba99,baj99,dm00}. 
Probability that a new link turns to be attached to a node is a function of its degree.  
Related ideas was discussed long ago by Simon \cite{s55}. All these models belong to a class of stochastic multiplicative processes \cite{stoch}. 

All the used models can be described by linear master equations \cite{dms00} which are solvable exactly in several particular cases. In this paper, we treat them using a trivial continuous approach \cite{baj99,dm00,dms00} that, as we show, describes main features of the networks and produces exact values of the exponents. Demonstration of possibilities of the continuous approach is one of the aims of our paper.

In our paper, we use several different models of evolving networks (with undirected and directed links), 
but a skeleton of all them is the same (see Fig. \ref{f1}). At each increment of time, a new node is added. Therefore, the number of the last node, $t$, may be called ``time''. Each node is labeled by the time of its birth, $s=0,1,2,\ldots \leq t$. Together with a new node, several new links is appeared in the network. A part of 
these links is distributed between nodes preferentially according to their degrees 
(we notate it $k$) or in-degrees ($\equiv q$). 
(Out-degree distributions is discussed in only one place).  
In general considerations, $k$ and $q$ will be equivalent if no special notion will be made.

The following ``one-node'' characteristics may be introduced. 
$p(k,s,t)$ is the probability that the node $s$ has degree $k$ at time $t$. 
$P(k,t)$ is total degree distribution of a network in time $t$. 
$P(k) = P(k,t \to \infty)$ is stationary distribution, $\gamma$ is the corresponding scaling exponent, $P(k) \propto k^{-\gamma}$ for large $k$. 
$\overline{k}(s,t)$ is the average degree of the node $s$ at time $t$. $\beta$ is its scaling exponent, $\overline{k}(s,t) \propto s^{-\beta}$. 
The corresponding definitions for in-degree distributions are similar.
   
In Sec. \ref{s-reasons} we introduce the continuous approach. Considering a simple example we 
explain reasons to use this approach and derive useful relations. 

In Sec. \ref{s-scaling1}, for linearly growing scale-free networks (total number of links is proportional to a total number of nodes with time-independent coefficient), using general considerations, we obtain scaling forms of involved quantities and 
universal relations between scaling exponents. 

Sec. \ref{s-types} is devoted to study of different types of preferential linking. 
Using the continuous approach, we describe (i) the simplest network in which new links are attached without any preference (produces exponential distributions), (ii) linear type of preferential linking (produces scale-free networks with $2<\gamma<\infty$), (iii) a mixture of preferential and random linking (also produces scale-free networks with $2 < \gamma < \infty$), and (iv) more general cases also producing scale-free networks. In fact, in Secs. \ref{s-types}, \ref{s-aging}, and \ref{s-condens} different realizations of a linear preference function, $G(s,t)k+A(s,t)$, according to which new links are distributed among nodes, are studied. In Sec. \ref{s-aging}, ``strength'' of a node, $G(s,t)$, depends only on 
its age, $G(s,t) \propto (t-s)^{-\delta}$. This changes scaling exponents. 
In Sec. \ref{s-condens}, strength of nodes depends only on dates of their birth, $s$. 
In this case, the network may exibit an intriguing phenomenon of capture of a finite fraction of all links by the strongest node -- giant capture, -- that was quite recently reported \cite{bb00b}. Nevertheless, we demonstrate that the network stays to be scale-free. 

Note that, in most of considered here models, new links may also connect old nodes. Nevertheless,  
in Sec. \ref{s-density}, we study a specific growing network in which new connections are possible only between a new node and old ones, that, in particular, may be used for description of networks of scientific citations. In this simple situation, in the continuous approach,  it is possible to describe distributions of links between pairs of individual nodes. 

Nonlinearly growing networks are considered in Sec. \ref{s-accelerating}. For this, more general case, we obtain scaling relations, and describe non-stationary distributions, $P(q,t)$, and their cut-offs impeding observation of power-law dependences for finite-size networks. 

In Sec. \ref{s-developing}, we study influence of {\em permanent damage} on the scaling characteristics of growing networks. We show that such type of the damage produces much stronger effect on the network than a previously studied instant damage \cite{ba00a,ceah00,cnsw00,ceah00a}. 

High quality of the used continuous approach is discussed in Sec. \ref{s-applicability} using already known exact results.

%%%%%%%%%%
%%%%%%%%%%
%%%%%%%%%%
%%%%%%%%%%

\section{Reasons 
for 
%to use 
the continuous approach}\label{s-reasons}

We start from one of the most simple models of growing networks with preferential linking proposed by Barab\'{a}si and Albert \cite{ba99} which belongs to classes of more general models wich were solved exactly afterwards \cite{krl00,dms00}. At each increment of time, a node is appeared. It connects to one of old nodes chosen with probability proportional to degree of this old node, i.e., to a total number of its connections. Using introducing in Sec. \ref{s-introduction} notations, it is possible to write immediately the following master equation for degree probabilities of individual nodes 
(see \cite{dms00}), 

\begin{equation}
p(k,s,t+1) = \frac{k-1}{2t} p(k-1,s,t) + \left( 1 - \frac{k}{2t} \right) p(k,s,t)
\,   
\label{2--1}
\end{equation} 
with, e.g., $t=1,2,3,\ldots$ and $s=0,1,2,\ldots,t$, i.e., at time $t=1$, 
a pair of nodes, $s=0,1$, connected by a link is present. Therefore, the initial condition is 
$p(k,s=0,1, t=1) = \delta_{k,1}$ and 
$p(k,t,t) = \delta_{k,1}$ is the boundary condition. Eq. (\ref{2--1}) may be rewritten in the 
form,

\begin{eqnarray}
2t[p(k,s,t+1) - & & p(k,s,t)] = 
\nonumber
\\[5pt]
& & 
(k-1) p(k-1,s,t) - k p(k,s,t)
\, .  
\label{2--2}
\end{eqnarray} 
Passing to the continuous limit in $t$ and $k$ we get

\begin{equation}
2t\frac{\partial p(k,s,t)}{\partial t} + \frac{\partial [k p(k,s,t)]}{\partial k} = 0
\,   
\label{2--3}
\end{equation} 

and

\begin{equation}
\frac{\partial [k p(k,s,t)]}{\partial \ln\sqrt{t}} + \frac{\partial [k p(k,s,t)]}{\partial\ln k} = 0
\, .  
\label{2--4}
\end{equation} 

The solution of Eq. (\ref{2--4}) is
$k p(k,s,t) = \delta(\ln k - \ln\sqrt{t/s} + const)$. The boundary condition is fulfilled if it is of the following form, 

\begin{equation}
p(k,s,t) = \delta(k - \sqrt{t/s})
\, .  
\label{2--5}
\end{equation} 
Therefore, we see that the transition to the continuous limit in the master equation leads to the $\delta$-function form of degree distributions of individual nodes. 

The main quantity of interest is total degree distribution of all the network,

\begin{equation}
P(k,t) = \frac{1}{t+1}\sum_{s=0}^{t}p(k,s,t)
\, .  
\label{2--6}
\end{equation} 
In the continuous approximation, stationary degree distribution is of the form,

\begin{equation}
P(k) = P(k,t\to \infty) = \lim_{t \to \infty}\frac{1}{t}\int_{0}^{t}ds\,p(k,s,t)
\, .  
\label{2--7}
\end{equation} 
Inserting the obtained expression for $p(k,s,t)$, Eq. (\ref{2--5}), into Eq. (\ref{2--7}), one gets the continuous approximation result for this model, 
$P(k) = 2/k^3$ \cite{ba99,baj99}. 

Another way to obtain this expression is to derive an equation for the total probability, 
$P(k,t)$. Applying $\sum_{s=0}^{t}$ to both sides of Eq. (\ref{2--1}) and passing to the 
$t \to \infty$ limit, one gets 

\begin{equation}
P(k) + \frac{1}{2}[kP(k) - (k-1)P(k-1)] = \delta_{k,1}
\, .  
\label{2--7a}
\end{equation}
In the continuous $k$ limit, this equation is of the form 
$P(k) + (1/2)d[kP(k)]/dk = 0$. Its solution is $P(k) \propto k^{-3}$. 

Often, it is more simple to proceed in a bit different way. 
Let us introduce average connectivity of an individual node:

\begin{equation}
\overline{k}(s,t) = \sum_{k=1}^{\infty} k p(k,s,t) = \int_0^\infty dk\, k p(k,s,t)
\, .  
\label{3--1}
\end{equation}
%%this is possible (pass to the continuous limit) if $k p(k,s,t)$ has no singularity at small $k$. 
Applying $\int_0^\infty dk k$ to Eq. (\ref{2--3}) and integrating its right part by parts, we get 

\begin{equation}
\frac{\partial \overline{k}(s,t)}{\partial t} = \frac{\overline{k}(s,t)}{\int_0^t du\, \overline{k}(u,t)}
\, .  
\label{2--9}
\end{equation}
Here, we used that, in the particular case of this model, the total number of links in the network, 
$\int_0^t ds \int_0^\infty dk\,k p(k,s,t)$, equals $2t$. Boundary condition for Eq. (\ref{2--9}) is 
$\overline{k}(t,t)=1$. 
It is usefull to use equations similar to Eq. (\ref{2--9}) but not to study each time 
corresponding master equations. 

The meaning of Eq. (\ref{2--9}) is quite obvious -- each new link is distributeded homogeneously among all nodes with account of the particular form of preference. At first, this approach has been called ``mean field'' \cite{ba99,baj99} but afterwards it was shown that it is simply continuous approximation \cite{dm00,dms00}. 

One should note that we pass to the continuous limit for both variables -- $k$ and $t$. 
The continuous limit for $t$ is not dangerous if we consider large enough networks and do not study some peculiarities of distributions related with particular form of initial conditions 
(for a very simple model of a scale free network, an exact solution was found for all values of $k$ and $t$ \cite{dms00c}). The continuous in $k$ approximation needs more care, so we discuss its quality along all the paper.

%%%%%%%%%%
%%%%%%%%%%
%%%%%%%%%%
%%%%%%%%%%

\section{Scaling relations for scale-free networks}\label{s-scaling1}
 
In the continuous approach, knowledge of average degree of nodes, $\overline{k}(s,t)$, 
%%that is the solution of Eq. (\ref{2--9}), 
lets to obtain the total degree distribution, $P(k,t)$. 
(Now we do not restrict ourselves by any particular model and our general results are also valid for in-degree, $q$, (out-degree) distributions. Nevertheless, in any case, $\overline{q}(s,t)$ is the solution of an equation similar to Eq. (\ref{2--9}).) 

Indeed, one can write 

\begin{eqnarray}
P(k,t) & = & \frac{1}{t}\int_0^t ds\, \delta(k-\overline{k}(s,t)) = 
\nonumber
\\[5pt] 
& &  -\frac{1}{t}\left( \frac{\partial \overline{k}(s,t)}{\partial s} \right)^{-1}\!\!\!\!\![s=s(k,t)]
\, ,  
\label{3--2}
\end{eqnarray}
where $s(k,t)$ is a solution of the equation, $k = \overline{k}(s,t)$.

Now, we can easily connect involved quantities. 
Assuming that $P(k)$ and $\overline{k}(s,t)$ exibit scaling behavior, that is 
$P(k) \propto k^{-\gamma}$ for large $k$ and $\overline{k}(s,t) \propto s^{-\beta}$ 
for $1 \ll s \ll t$, one gets $s \propto k^{-1/\beta}$ and 
$k^{-\gamma} \propto \partial s/\partial k \propto k^{-1-1/\beta}$. Therefore, 
$\gamma = 1 + 1/\beta$, and we obtain the following scaling relation:

\begin{equation}
\beta(\gamma-1) = 1
\, .  
\label{3--3}
\end{equation}

Let us show that the relation, Eq. (\ref{3--3}), is universal. Here, we can proceed using general considerations. In the present section, we study only linearly growing networks (input flow of new links does not depend on time) which produce stationary degree distributions at long times. More complex cases will be considered in Sec. \ref{ss-scaling2}. 

Hence, let $P(k)$ be stationary, then it follows from Eq. (\ref{2--7}) that $p(k,s,t)$ has to be of the form, 
$p(k,s,t)=\rho(k,s/t)$.
The normalization condition is 

\begin{equation}
\int_0^\infty dk\,p(k,s,t) = 1
\, ,  
\label{3--4}
\end{equation}
so $\int_0^\infty dk\,\rho(k,x) = 1$. From this equation, it follows that $\rho(k,x) = g(x) f(kg(x))$, where $g(x)$ and $f(x)$ are arbitrary functions. 
From the definition of $\overline{k}(s,t)$, Eq. (\ref{3--1}), using its scaling behavior, we get 
$\int_0^\infty dk\,k\rho(k,x) \propto x^{-\beta}$. Substituting $\rho(k,x)$ into this relation, 
one obtains that $g(x) \propto x^\beta$. Of course, without loss of generality, one may set $g(x)=x^\beta$, so we obtain the following scaling form of degree distributions of individual nodes,

\begin{equation}
p(k,s,t)=(s/t)^\beta f(k(s/t)^\beta) 
\, .  
\label{3--5}
\end{equation}
Finally, assuming scaling behavior of $P(k)$, i.e., $\int_0^\infty dx\, \rho(k,x) \propto k^{-\gamma}$, 
and using Eq. (\ref{3--5}), we obtain $\gamma=1+1/\beta$, i.e., the relation, Eq. (\ref{3--3}) is universal. 
Here we used quick convergence of $\rho(k,x)$ at large $x$ (it follows from our exact results 
\cite{dms00}).
One should note that, during this derivation, we did not use any approximations.

%%%%%%%%%%
%%%%%%%%%%
%%%%%%%%%%
%%%%%%%%%%

\section{Types of preferential linking 
%producing scale-free networks
}\label{s-types}

%%%%%
%%%%%%%%
%%%%%

\subsection{Absence of preference}\label{ss-absence} 

We need an example of a non-scale-free network for comparison, so  
we start with the simplest growing network with random attaching of new links. 
Let again, as in Sec. \ref{s-reasons}, at each increment of time, a new node be added to the network.
Now it connects with a randomly chosen (i.e., without any preference) old node. The growth begins 
from the configuration consisting of two connected nodes at time $t=1$. This model is even simpler than the well-known Erd\"{o}s-R\'{e}nyi model \cite{er60,b-book}, and its exact solution is trivial.
%$s=0,1,2,\ldots,t$, $t=1,2,3,\dots$ 
The master equation describing evolution of degree distributions of individual nodes looks as

\begin{equation}
p(k,s,t+1) = \frac{1}{t+1} p(k-1,s,t) + \left( 1 - \frac{1}{t+1} \right) p(k,s,t)
\, ,  
\label{4--1}
\end{equation}
$p(k,s=0,1,t=1) = \delta_{k,1}$, $\delta(k,t,t\geq 1) = \delta_{k,1}$. 
Applying $\sum_{s=0}^{t}$ to both sides of Eq. (\ref{4--1}) and using the definition of total degree distribution, Eq. (\ref{2--6}), 
%%$P(k,t)=\sum_{s=1}^t p(k,s,t) /t$, so 
we get

\begin{equation}
(t+1)P(k,t+1) - tP(k,t) = P(k-1,t) - P(k,t) + \delta_{k,1}
\, .  
\label{4--2}
\end{equation}
The corresponding stationary equation,

\begin{equation}
2P(k) - P(k-1) = \delta_{k,1}
\, ,  
\label{4--3}
\end{equation} 
has a solution of an exponential form, $P(k) = 2^{-k}$. Therefore, networks of such type often are called ``exponential''.

From Eq. (\ref{4--1}), one may also find the degree
distribution of individual nodes, $p(k,s,t)$, for large $s$ and $t$ and fixed $s/t$,

\begin{equation}
p(k,s,t) = \frac{s}{t} \frac{1}{(k+1)!} \ln^{k+1}\left(\frac{t}{s}\right)
\, .  
\label{4--4}
\end{equation}
One sees that this function decreases rapidly at large values of degree, $k$. 

Thus, degree distributions of networks growing without using a mechanism of preferential linking differs strikingly from distributions of scale-free networks described in previous section. 
Note that the singularity of $p(k,s,t)$, Eq. (\ref{4--4}), at $s/t \to 0$ is much weaker than for scale-free networks.

%%%%%
%%%%%%%%
%%%%%

\subsection{Linear preference}\label{ss-linear1}

\begin{figure}
\epsfxsize=85mm
\epsffile{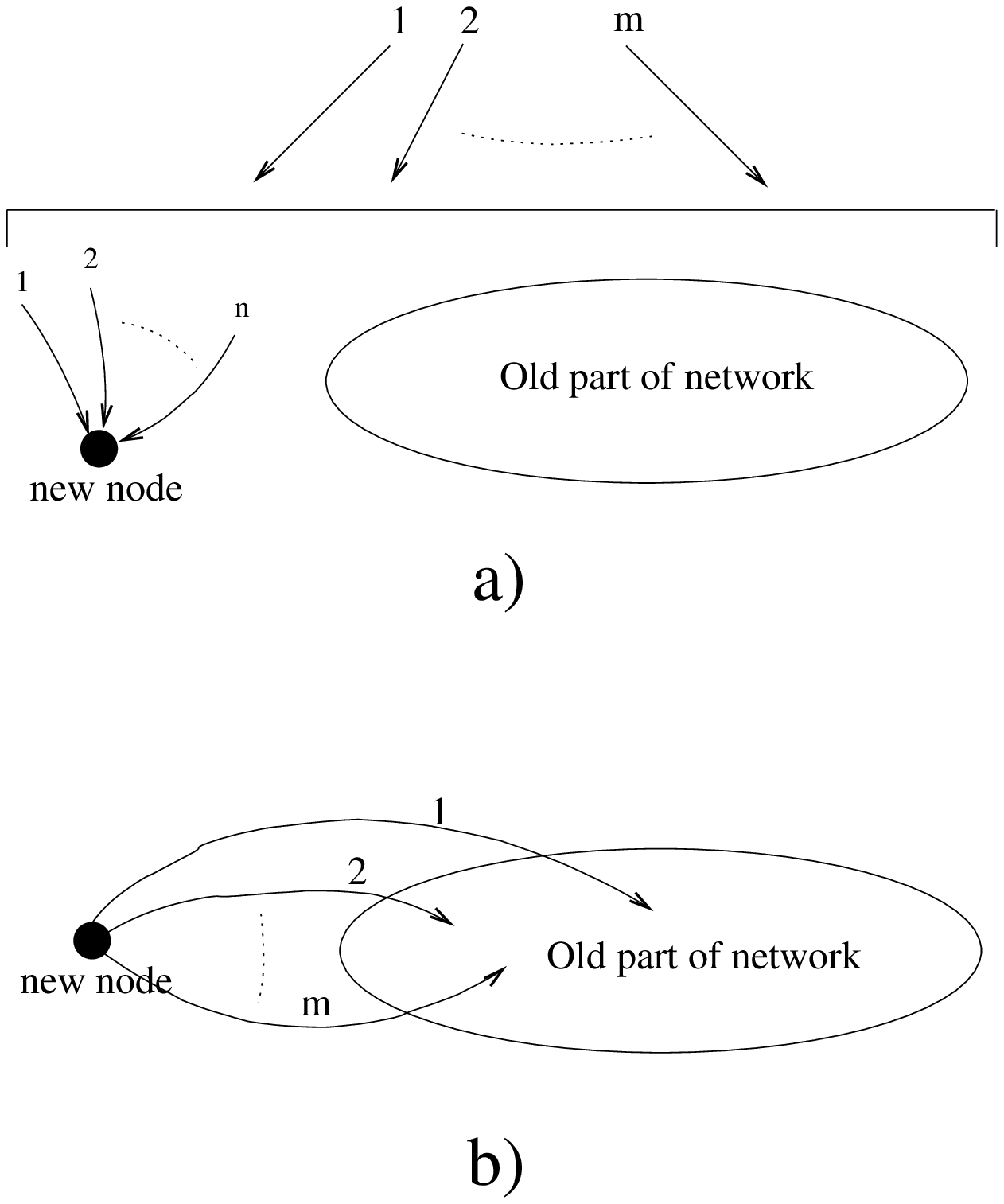}
\caption{
Scheme of growth of two standart network models considered in the present paper. 
(a) At each time step, a new node with $n$ incoming links is added. Source
ends of all links are placed anywhere. In addition, $m$ links are distributed preferentially among all nodes. This means that a target end of each of these links is attached to some node according to a rule of preference. 
(b) A citation network model. $n=0$. At each increment of time, a new node with $m$ 
outgoing links is added. Target ends of these links are distributed preferentially among old nodes.
}
\label{f1}
\end{figure}

Let us demonstrate the continuous approach in details using a more general model than in Sec. \ref{s-reasons} 
which produce a wide range of $\gamma$ exponent values \cite{dms00} (see Fig. \ref{f1},\,a). Let us consider a network with directed links. Here, we study distributions of numbers of {\em incoming} links of nodes ({\em in-degree}, $q$). 

At each time step a new node is added. It has $n$ incoming links. These links go out from arbitrary nodes or even from some external source. Simultaneously, $m$ extra links are distributed with preference. This means that again they go out from non-specified nodes or from an external source but a target end 
of each of them is attached to a node chosen preferentially: probability to chose some particular node is proportional to a function of its in-degree, we call it a {\em preference function}. In the simple model that we consider in the present section, this probability is proportional to $q+A$, where $A$ is a constant which we call {\em additional attractiveness}. We shall see that its reasonable 
values are $A>-n$. 

In the continuous approach, we may assume that $m$ and $n$ are not necessary integer numbers but any positive ones. 
We do not worry about source ends of links because study here only in-degree distributions.      

Then the equation for average in-degree is of the form, 

\begin{equation}
\frac{\partial \overline{q}(s,t)}{\partial t} = m \frac{\overline{q}(s,t)+A}{\int_0^t du [\overline{q}(u,t) + A]}
\,   
\label{5--1}
\end{equation}
with the initial condition, $\overline{q}(0,0)=0$, and the boundary one, $\overline{q}(t,t)=n$.

Applying $\int_0^t ds$ to Eq. (\ref{5--1}) we obtain

\begin{equation}
\frac{\partial}{\partial t} \int_0^t ds\,\overline{q}(s,t) = 
\int_0^t ds\,\frac{\partial}{\partial t} \overline{q}(s,t) + \overline{q}(t,t)
\, , 
\label{5--2}
\end{equation}
so 

\begin{equation}
\int_0^t ds\,\overline{q}(s,t) = (m+n)t
\, . 
\label{5--3}
\end{equation}
This relation is quite natural -- the total in-degree of the network is equal to the total number of created links, and we see that Eq. (\ref{5--1}) is consistent.
Therefore, Eq. (\ref{5--1}) takes the form, 

\begin{equation}
\frac{\partial \overline{q}(s,t)}{\partial t} = 
\frac{m}{m+n+A} \frac{\overline{q}(s,t)+A}{t}
\, .  
\label{5--4}
\end{equation}
Its general solution is 

\begin{equation}
\overline{q}(s,t)+A = C(s) t^{m/(m+n+A)}
\, .  
\label{5--5}
\end{equation}
where $C(s)$ is arbitrary function of $s$. 
Accounting for the boundary condition, $\overline{q}(t,t)=n$, one has 

\begin{equation}
\overline{q}(s,t)+A = (n+A) \left(\frac{s}{t}\right)^{-m/(m+n+A)}
\, .  
\label{5--6}
\end{equation}
Hence, the scaling exponents are 

\begin{equation}
\beta = \frac{m}{m+n+A}
\,   
\label{5--7} 
\end{equation}
and 

\begin{equation}
\gamma = 2 + \frac{n+A}{m}
\, .  
\label{5--8} 
\end{equation}
One sees that, for $n+A>0$, the exponent $\gamma$ is in the range $2<\gamma<\infty$ while $\beta$ 
belongs to the interval, $0<\beta<1$. We see that $n$ plays the same role as $A$.

If we set $n=0$ and demand that all new links have to go out from new nodes, we get a network of citations (see Fig. \ref{f1},\,b).
Note that, when $A=m$ and $n=0$, we get the particular case of the Barab\'{a}si-Albert's model 
\cite{ba99} in which each new link is connected with a new node. Indeed, in this case, degree and in-degree of nodes are coupled rigidly, $k=q+m$, and the obtained results are valid for the degree distributions. If, in addition, we set $m=1$, we obtain the model considered in Sec. \ref{s-reasons}.

%%%%%
%%%%%%%%
%%%%%

\subsection{Mixture of preferential and random linking}\label{ss-mixture}

Now we can consider a bit more complex model. We will demonstrate that scale-free networks may be obtained even without ``pure'' preference linking. We discuss the model introduced in the previous section but with one new element. 
Let, in addition, at each time step, $n_r$ links be distributed between nodes randomly, without any preference. Again, we study in-degree distributions, so these links may go out from anywhere but their target ends are attached to randomly chosen nodes. 
Then, 

\begin{equation}
\frac{\partial \overline{q}(s,t)}{\partial t} = 
\frac{n_r}{t} + m \frac{\overline{q}(s,t)+A}{\int_0^t du [\overline{q}(u,t) + A]}
\, .  
\label{6--1}
\end{equation}
Again, $\overline{q}(0,0)=0$ and $\overline{q}(t,t)=n$. In this case, 
$\int_0^t ds\,\overline{q}(s,t) = (n_r+m+n)t$, so Eq. (\ref{6--1}) takes the form:

\begin{eqnarray}
\frac{\partial \overline{q}(s,t)}{\partial t} = & & \frac{m}{n_r+m+n+A}\times
\nonumber
\\[5pt]
& & 
\frac{1}{t} \left[\overline{q}(s,t) + A + \frac{n_r}{m}(n_r+m+n+A) \right]    
\label{6--2}
\end{eqnarray}
Its solution, with account for the boundary condition is 

\begin{eqnarray}
& & \overline{q}(s,t) + A + \frac{n_r}{m}(n_r+m+n+A) = 
\nonumber
\\[5pt]
& & \frac{n_r+m}{m}(n+n_r+A)\left(\frac{s}{t}\right)^{-m/(n_r+m+n+A)}
\, .  
\label{6--3}
\end{eqnarray}
Here, $m$, $n$, and $n_r$ are positive numbers, and $n_r+n+A>0$. 

Therefore, we get the scale-free network with the exponents, $\beta = m/(m+n_r+n+A)$ and

\begin{equation}
\gamma = 2 + \frac{n_r+n+A}{m}
\, .  
\label{6--4} 
\end{equation} 
Thus, additional fraction of randomly distributed links does not delete a power-law dependence of distributions but only increase $\gamma$.

In two computer-science papers \cite{kkrrt99,krrt99} a similar model of a network with directed links was considered.
At each time step, a new node is added to the net. 
It has one outgoing link.
The other end of this link connects with one of old nodes by the following rule.
(i) With probability $p$ it connects with a random node. 
(ii) With probability $1-p$ it finds a random node and connects with its sole target neighbor node. 
(In this particular case, it is the same as to chose random link and to connect with its target end.)

One can see that this model corresponds to the particular case, $n=0, n_r=p, m=1-p, A=0$, 
of the network that we consider in the present Section. From Eq. (\ref{6--4}), we get 
$\gamma= 2+ p/(1-p) = 1 + 1/(1-p)$. 
(Note that, in these papers Refs. \cite{kkrrt99,krrt99}, unproper result figurates. The necessary additional unit is absent there. Indeed, if one of the factors (random linking) produces $\gamma=\infty$, another one (preferential linking) -- $2<\gamma<\infty$, then their interplay can not produce $\gamma<2$.) 

It is worth while to emphasize here a close connection of the models that we consider with the well-known Simon's model \cite{s55} discussed in relation with networks in \cite{be00}. 
In the Simon's model, if one uses the language of growing networks with directed links, 
at each time step, a new link appears. Since we discuss here only in-degree, it is again not important, from were it goes out. With some fixed probability, a new node is created and the 
target end of the link is attached to it. With an adjusted probability, the target end of the link is attached to the target end of a randomly chosen old nodes. Then only one point in the Simon's model, differs from the models that we consider in the present paper: at each increment of time, a link is added but not the node as in our case. Of course, this does not influence results for large networks and scaling exponents.

%%%%%
%%%%%%%%
%%%%%

\subsection{WWW exponents}\label{ss-www}

The previous result for the $\gamma$ exponent, Eq. (\ref{6--4}), lets us to obtain the most rude estimates for the exponents of the World Wide Web \cite{be00,dms00b} which are measured with sufficient precision for comparison. Let us apply the model of Sec. \ref{ss-mixture} to the growth of the World Wide Web. This means that we have to assume that each time when a new node of the Web appears, in average, the same number of new links arises between its nodes. We neglect many very important factors including eliminating of some nodes and links during the growth, etc. 

We do not know values of any of the quantities in the left part of Eq. (\ref{6--4}). 
The constant $A$ may takes {\em any} values between $-(n_r+n)$ and infinity, the number of the randomly distributed links, $n_r$, in principle, may be not small (there exist many persons making their references practically at random), $n$ is not fixed. From the experimental data \cite{bkm00}, we know more or less the sum, $m+n+n_r \sim 10 \gg 1$ (between 7 and 10, more precisely), and that is all. 

The only thing we can do, it is to fix the scales of the quantities. The natural characteristic values for $n_r+n+A$ in Eq. (\ref{6--4}) are (a) $0$, (b) $1$, (c) $m \gg 1$, and (d) infinity. In the first case, 
each node have zero initial attractiveness, and all new links are directed to the oldest node, $\gamma \to 2$. In the last case, there is no preferential linking, and the network is not scale-free, $\gamma \to \infty$. Let us consider the really important cases (b) and (c). 

{\bf (b)} \ How do pages appear in the Web? Suppose, you want create your personal home page. Of course, first you prepare it, put references, etc. But that is only the first step. You have to make it accessible in the Web, to launch it. You come to your system administrator, he put a reference to it (usually one reference) in the home page of your institution, and that is more or less all -- your page is in the World Wide Web. 
There is another way of appearing of new documents in the Web. Imagine that you already have your personal home page and want to launch a new document. The process is even simpler than the one described above. You simply insert at least one reference to the document into your page, and that is enough for the document to be included in the World Wide Web.  
If the process of appearing of each document in the Web is as simple as the creating of your page -- only one reference to the new document $(n=1)$ -- and if one forgets about the terms $n_r$ and $A$ in  Eq. (\ref{6--4}), 
than, for the $\gamma_{in}$ exponent of the distribution of the incoming links (in-degree distribution), we get immediately the estimation, $\gamma_{in}-2 \sim 1/m \sim 10^{-1}$. 
This estimation is indeed coincides with experimental value $\gamma_{in}-2=0.1$ \cite{bkm00}. 
(Here, we have introduced the new notation, $\gamma_{in}$, because in the present section we discuss different distributions.) Therefore, the estimation seems to be good. 
Nevertheless, we should repeat again, that this estimation follows only from the fixation of scales of the involved quantities. 
We emphasize that there are no any general reasons to set, e.g., $A=0$. 
A lot of real processes are not included in this estimation. Aging of nodes changes $\gamma$ 
see Sec. \ref{s-aging} and \cite{dm00}, account for dying of nodes $\gamma$, see \cite{dm00b} and  (the half-life of a page in the Web is of the order of half a year) changes $\gamma$. The ratio between the total number of links and the number of nodes in the Web is not constant \cite{bkm00}, it increases with time, the growth of the Web is nonlinear. This factor also change the value of $\gamma$, in future it may become even lower than $2$, see \cite{dm00c} and Section \ref{s-accelerating}.  

{\bf (c)} \ Above we discussed the distribution of incoming links. Eq. (\ref{6--4}) may be also applied for the distribution of links which come out from documents of the Web, since the model of the previous section may be reformulated for outgoing links of nodes.  
In this case all the quantities in Eq. (\ref{6--4}) takes other values which are again unknown. Nevertheless, one may think that the number of the links distributing without any preference, $n_r$, is not small now. Indeed, even beginners proceed with linking of their pages. 
Hence, $n_r \sim m$ -- we have no another available scale, -- and $\gamma_{out}-2 \sim m/m \sim 1$. 
We again can compare this estimation with the experimental value, $\gamma_{out}-2=0.7$ \cite{bkm00}.

%%%%%
%%%%%%%%
%%%%%

\subsection{Generalized form of preference}\label{ss-generalized} 

What are the other forms of preference that produce scale-free networks? Let us list the main possibilities. 
In the present section and in 
Secs. \ref{s-aging} and \ref{s-condens} we consider only a linear form of a preference function. Nonlinear preference functions are discussed in Sec. \ref{s-applicability}.  

The reasonable (linear) form of the probability for a new link be attached to a node $s$ at time $t$ is 
$p_{s,t} = G_{s,t} k_{s,t} + A_{s,t}$. The coefficient $G_{s,t}$ may be called {\em strength} of a node  
(in Refs. \cite{bb00b,bb00a} it is called ``fitness'') $A_{s,t}$ is {\em additional attractiveness}. 

One can consider the following particular cases.

(i) $G=const$, $A=A(s)$. 
In this case, the additional attractiveness, $A(s)$ may be treated as ascribed to individual nodes. 
A possible generalization is to make it a random quantity. One can check that the answers do not change crucially -- one only has to substitute the average value, $\overline{A}$, instead of $A$, into previous 
expressions for scaling exponents.

Note that $n$ and $m$ may be maked also random and substituted by $\overline{n}$ and $\overline{m}$ in the expressions for the exponents. 

There is a more interesting possibility -- to construct a direct generalization of the network considered in Sec. \ref{ss-mixture}. 
For this, we may ascribe the additional attractiveness not to nodes but to new links and again make it random quantity. Therefore, new links play the role of fans with different passion for popularity of their idols, nodes. This is case (ii), $G=const$, $A=A(t)$, where $A(t)$ is random. 

Let a distribution function of $A$ be $P(A)$. Then, our main equation looks as

\begin{equation}
\frac{\partial \overline{q}(s,t)}{\partial t} = 
\overline{m} \int dA\,P(A)\,\frac{\overline{q}(s,t)+A}{\int_0^t du [\overline{q}(u,t) + A]}
\, .  
\label{8--1}
\end{equation}
Initial and boundary conditions are $\overline{q}(0,0)=0$ and $\overline{q}(t,t)=\overline{n}$. 
Hence, $\int_0^t ds\, \overline{q}(s,t) = (\overline{n} + \overline{m})t$. 
Therefore,

\begin{equation}
\frac{\partial \overline{q}(s,t)}{\partial t} = 
\frac{\overline{m}}{t}
\left[ 
\overline{q}(s,t)
\int \frac{dA\,P(A)}{\overline{n} + \overline{m} + A} +
\int \frac{dA\,A P(A)}{\overline{n} + \overline{m} + A}
\right]
\, .  
\label{8--2}
\end{equation}
Hence we again obtain a scale-free network, 

\begin{equation}
\beta = \overline{m} \int \frac{dA\,P(A)}{\overline{n} + \overline{m} + A}
\,   
\label{8--3} 
\end{equation}
and 

\begin{equation}
\gamma = 1 + 
\left[ 
\int \frac{dA\,P(A)}{1 + (\overline{n} + A)/\overline{m}} 
\right]^{-1}
\, .  
\label{8--4} 
\end{equation}
These expressions generalize the corresponding results of Secs. \ref{ss-linear1} and \ref{ss-mixture}. 
One can see that the values of the $\gamma$ exponent are again between $2$ and $\infty$, $\beta$ is between $0$ and $1$.  

(iii) $A=const$.  

The case, $G(s,t) \propto (s-t)^{-\Delta}$ is considered in Sec. \ref{s-aging}, it produces $2<\gamma<\infty$. 

Different versions of the $s$-dependent strength, $G(s,t)=G(s)$, are considered in Sec. \ref{s-condens}. 
The case of the power-law dependence, $G(s) \propto s^{-\Delta^\prime}$ 
is considered in Sec. \ref{ss-varying} -- three different types of behavior are possible: 
$\Delta^\prime<0$ -- an exponential network; $\Delta^\prime=0$ -- scale-free; $\Delta^\prime>0$ -- the oldest 
node gets a finite part of total degree of a network. 
%plus weak long tail $P(k) \propto 1/(q\ln^{1+1/\Delta}q)$. 

One can consider a network in which several nodes are ``stronger'' then others. We investigate effects arising in this situation in Sec. \ref{ss-local}. 
A homogeneous mixture of nodes with two different strengths is studied in 
Sec. \ref{ss-mix}.  

The case of fluctuating $G(s)$ is considered by Bianconi and Barab\'{a}si \cite{bb00a} -- 
when the distribution of $G$ is homogeneous, i.e., $G$ is homogeneously distributed between two fixed values,
the distribution has a logarithmic factor, 
$ P(q) \propto q^{-\gamma}/\ln q $, $2<\gamma<\infty$.  
For some special forms of the distribution of $G$, strong cooperative effects were found recently \cite{bb00b}.

(iv) $A=const$, $G=G(t)$. This case is reduced to the case (ii).

%%%%%%%%%%
%%%%%%%%%%
%%%%%%%%%%
%%%%%%%%%% 

\section{Effect of aging of nodes}\label{s-aging}

How does the structure of the network change if one introduces aging of sites \cite{dm00}, i.e., if 
the probability of connection of the new site with some old one is proportional not only to the connectivity of the old site but also to the power of its age, $\tau^{-\Delta}$, for example? Here, we introduce the aging exponent, $\Delta$. Such aging is natural for networks of scientific citations in which old papers usually have low attractiveness. 

\begin{figure}
\epsfxsize=70mm
\epsffile{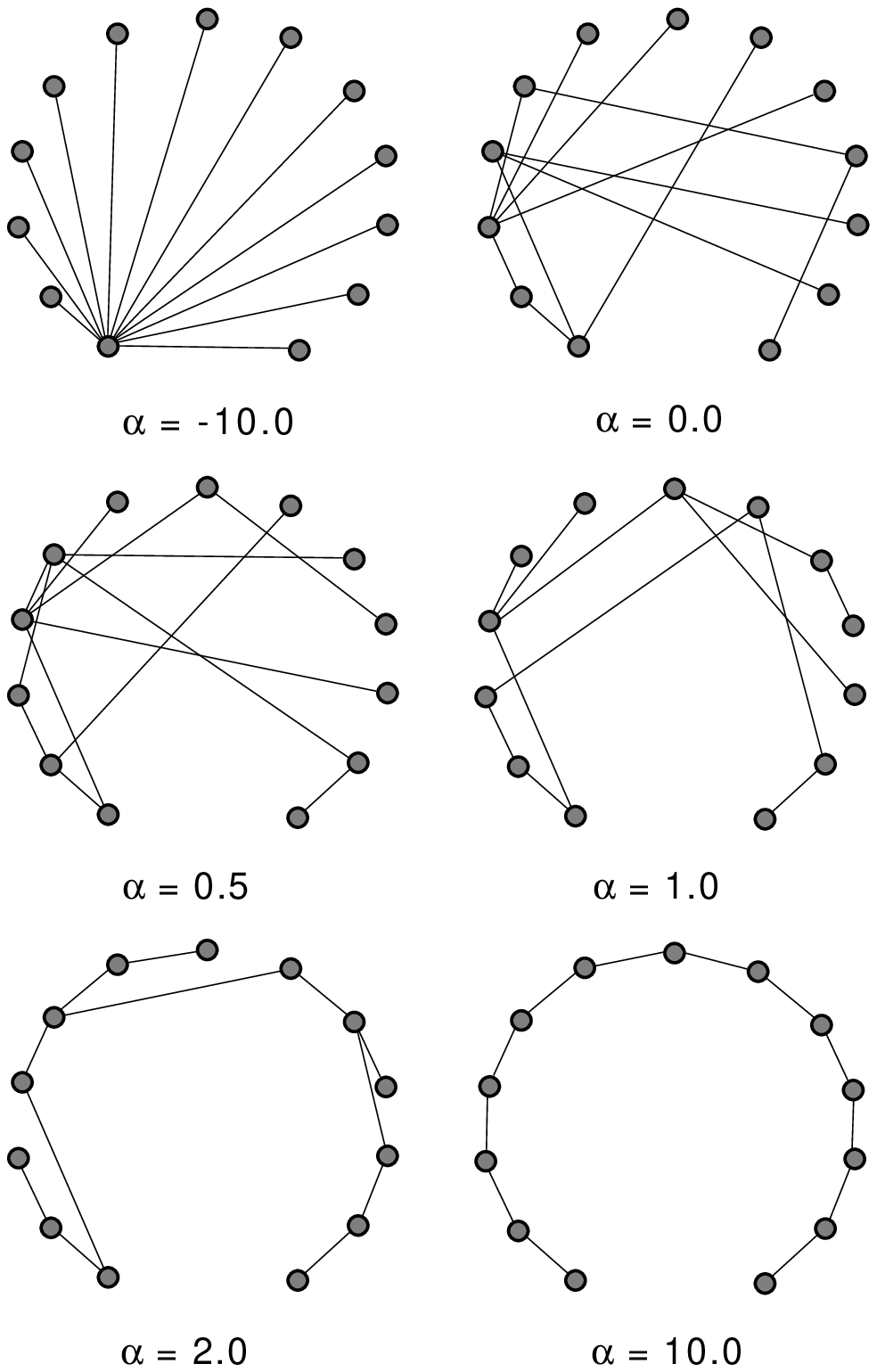}
\caption{
Change of the structure of the 
network with aging of nodes with increase of the aging exponent $\alpha$. 
The aging is proportional to $\tau^{-\alpha}$, where $\tau$ is an age of the site. 
The network grows clockwise starting from the site below on the left. Each time 
one new site with one link is added.
}
\label{f2}
\end{figure}

The simplest analytical expressions may be obtained for the model of Sec. \ref{s-reasons}: links are undirected, and every new node is connected by a single link with some old node which is chosen according to the same rule of preference as in Sec. \ref{s-reasons}. Fig. \ref{f2} demonstrates that the structure of the network depends crucially on the value of the aging exponent. When $\Delta$ is large enough (larger than $1$), the network turns to be a linear structure. For low negative values of $\Delta$, i.e., for large $-\Delta$ (attractiveness of documents increase with growth of their age) many links are attached to the oldest nodes. 

The resulting equation for such a network is of the form, 

\begin{equation}
\frac{\partial \overline{k}(s,t)}{\partial t} = 
\frac{\overline{k}(s,t)(t-s)^{-\Delta}}{\int\limits_0^t du \overline{k}(u,t)(t-u)^{-\Delta}} \
,  \ \overline{k}(t,t)  =  1 
\label{9--1}
\, .
\end{equation}
%(only one link is added each time). 
Then, $\int_ 0^t ds \overline{k}(s,t) = 2t$.

We search for the solution of Eq. (\ref{9--1}) in the scaling form,
\begin{equation}
\overline{k}(s,t) \equiv \kappa(s/t) \ , \ s/t \equiv \xi 
\label{9--2}
, .
\end{equation}
Then  Eq. (\ref{9--1}) becomes 

\begin{eqnarray}
-\xi (1-\xi)^\Delta \frac{d \ln \kappa(\xi)}{d \xi}  & = & 
\left[ \int_0^1 d\zeta \kappa(\zeta) (1-\zeta)^{-\Delta} \right]^{-1} \equiv \beta 
, 
\nonumber \\[3pt]
\kappa(1) = 1
\, ,
\label{9--3}
\end{eqnarray}
where $\beta$ is a constant which is unknown yet. One can understand that it is just the exponent $\beta$ of average connectivity of individual nodes since, 
in the scaling region, $\xi \ll 1$, Eq. (\ref{9--3}) provides the power law dependence 
$\kappa(\xi) \propto \xi^{-\beta}$. We also get the relation $\int_0^1 d\zeta \kappa(\zeta) = 2$. 
Our aim is to find $\beta$. 
For this, we have to find a solution of Eq. (\ref{9--3}) containing this unknown parameter $\beta$ and substitute it into its definition, the left part of Eq. (\ref{9--3}), or into the relation for total number of links.

The solution of Eq. (\ref{9--3}) is 
\begin{equation}
\kappa(\xi) = B \exp \left[-\beta \int \frac{d\xi}{\xi(1-\xi)^\Delta}  \right]
\, ,
\label{9--4}
\end{equation}
where $B$ is a constant. The indefinite integral in Eq. (\ref{9--4}) may be taken:

\begin{eqnarray}
\label{8}
& & \int \frac{d\xi}{\xi(1-\xi)^\Delta} = 
\nonumber \\
& & \ln \xi + \sum_{j=0}^{\infty} \frac{1}{j!(j+1)^2} 
\Delta(\Delta+1) \ldots (\Delta+j)\xi^{j+1}  =
\nonumber 
\\ [5pt]
& & \ln \xi + \Delta\, _3 F_2(1,1,1+\Delta;2,2;\xi)
,
\label{9--5}
\end{eqnarray}
where $_3 F_2(\ ,\ ,\ ;\ ,\ ;\ )$ is the hypergeometric function.
Recalling the boundary condition $\kappa(1) = 1$, we find the constant $B$. Thus the solution is

\begin{eqnarray}
& & \kappa(\xi) = 
\nonumber 
\\[5pt]
& & e^{-\beta[C + \psi(1-\Delta)]} \xi^{-\beta} \exp 
\left[ -\beta \Delta \xi\, _3 F_2(1,1,1+\Delta;2,2;\xi)  \right]
\, ,
\label{9--6}
\end{eqnarray}
where $C=0.5772 \ldots$ is the Euler's constant and $\psi(\ )$ is the $\psi$-function. 
%%Now we see that the constant $\beta$ indeed is the exponent of mean connectivity, 
%%since $\kappa(\xi) \sim \xi^{-\beta}$ if $\xi \to 0$. 
The transcendental equation for $\beta$ may be written 
if one substitutes 
Eq. (\ref{9--6}) into the right side of Eq. (\ref{9--3}):

\begin{eqnarray}
\beta^{-1} =  e^{-\beta[C + \psi(1-\Delta)]} \times     \phantom{WWWWWWWWWWW}
\nonumber 
\\[5pt]
\int_0^1 \frac{d\zeta}{\zeta^{\beta} (1-\zeta)^{\Delta}} \exp 
\left[ -\beta \Delta \zeta\, _3 F_2(1,1,1+\Delta;2,2;\zeta)  \right]
\, .
\label{9--7}
\end{eqnarray}
(An equivalent equation one may obtain substituting the solution, 
Eq. (\ref{9--6}), into the relation, $\int_0^1 d\zeta \kappa(\zeta) = 2$). 
The $\gamma$ exponent may be obtained using the universal relation, Eq. (\ref{3--3}).  

 The solution of Eq. (\ref{9--7}) exists in the range, $-\infty < \Delta < 1$. One may also find 
simple expressions for $\beta(\Delta)$ and $\gamma(\Delta)$ 
at $\Delta \to 0$: 

\begin{equation}
\beta \cong \frac{1}{2} - (1 - \ln 2)\Delta \ , \ 
\gamma \cong 3 + 4(1 - \ln 2)\Delta
\, ,
\label{9--8}
\end{equation}
where the numerical values of the coefficients are 
$1 - \ln 2 = 0.3069\ldots$ and $4(1 - \ln 2) = 1.2274\ldots$.
We used the relation
$_3F_2(1,1,1;2,2;\zeta) = Li_2(\zeta)/\zeta \equiv 
(\sum_{k=1}^\infty \zeta^k/k^2)/\zeta$
while deriving Eq. (\ref{9--8}). Here, $Li_2(\ )$ is the polylogarithm function of order 2.

\begin{figure}
\epsfxsize=75mm
\epsffile{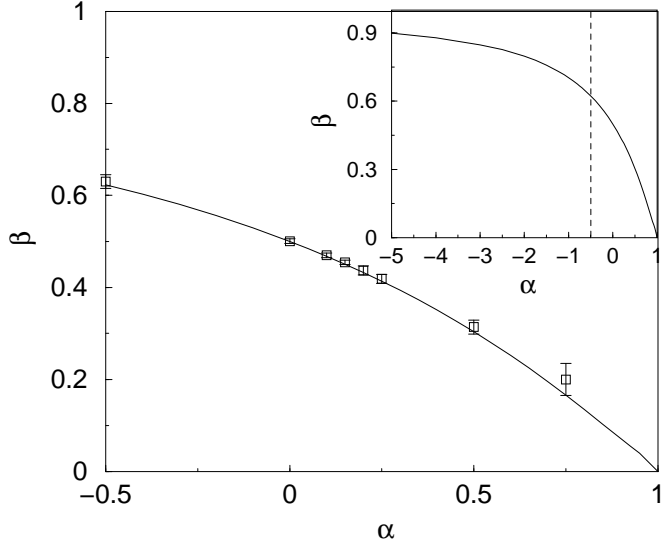}
\caption{
Exponent  $\beta$ of the mean connectivity vs. aging exponent $\alpha$. Points are obtained from simulations \protect\cite{xxx}. The line is the solution 
of Eq. (\protect\ref{9--7}). The inset shows the analytical solution in the range $-5<\alpha<1$. 
Note that $\beta \to 1$ if $\alpha \to -\infty$.
}
\label{f3}
\end{figure}

\begin{figure}
\epsfxsize=75mm
\epsffile{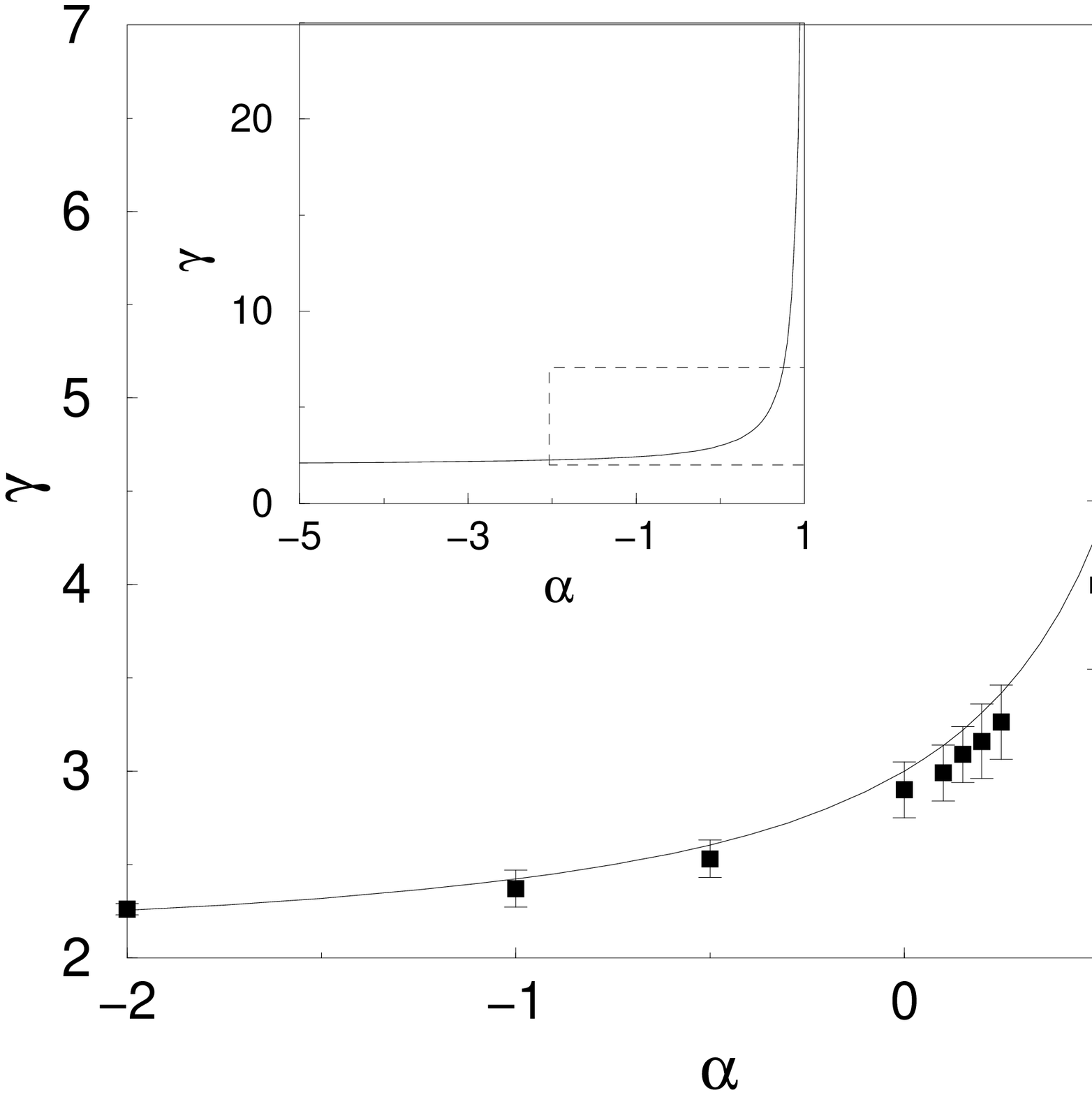}
\caption{
Exponent  $\gamma$ of the connectivity distribution vs. aging exponent $\alpha$. The points show the results of simulations \protect\cite{xxx}. The line is the solution of 
Eq. (\protect\ref{9--7}) with account for Eq. (\protect\ref{3--3}). The inset depicts the analytical solution in the range $-5<\alpha<1$.
}
\label{f4}
\end{figure}

In the limit of $\Delta \to 1$, using the relation
$_3F_2(1,1,2;2,2;\zeta) = -\ln(1-\zeta)/\zeta$, we find

\begin{equation}
\beta \cong c_1 (1-\Delta) \ , \ 
\gamma \cong \frac{1}{c_1} \frac{1}{1-\Delta}
\, .
\label{9--9}
\end{equation}
Here,
$c_1=0.8065\ldots,c_1^{-1}=1.2400\ldots$: the constant $c_1$ is the solution of the equation
$1+1/c_1 = \exp(c_1)$. 
The dependences, $\beta(\Delta)$ and $\gamma(\Delta)$, are shown in Figs. \ref{f3} and \ref{f4}. 
For comparison, data of simulations \cite{dm00} are also plotted.  
One sees, that $\beta \to 1$ and $\gamma \to 2$ 
in the limit $\Delta \to -\infty$. Therefore, the whole range of the variation of $\beta$ is $(0,1)$ 
and of $\gamma$ is $(2,\infty)$.  
One should note that, in this section, we studied the case of $m=1$ but the results, the scaling exponents are, of course, independent of $m$.

%%%%%%%%%%
%%%%%%%%%%
%%%%%%%%%%
%%%%%%%%%%

\section{Capture of links by ``strong'' nodes}\label{s-condens}

Now we can consider effects of $s$-depending strength of nodes, $G=G(s)$. 
We proceed with our tactics and demonstrate these effects using minimal models. 

It is worth to start this section from the following notion. All networks that we study in the present paper have a general feature -- {\em each} their node has a chance to get a new link. Only one circumstance prevent their enrichment -- seizure of this link by another node. In such kinetics of distribution of links, there are no any finite radius of ``interaction'' and there are no principal obstacles for capture of great fraction of links by some node. 

%%%%%%
%%%%%%
%%%%%%

\subsection{Varying strength of nodes}\label{ss-varying} 

Let us start from the case of a power-law dependence of the strength, 
$G(s) \propto s^{-\Delta^\prime}$. At first sight, this problem seems to be very close to the one 
that was considered in the previous section. Nevertheless, repetition of calculations of 
Sec. \ref{s-aging} for the present model leads immediately to the following expression for the degree, 
 
\begin{equation}
\kappa(\xi) = B \exp [\beta \xi^{-\Delta^\prime}]
\, ,
\label{18--1}
\end{equation}
where $B$ is a constant, $\xi=s/t$, $\overline{k} = \kappa(\xi)$ 
(compare with Eq. (\ref{9--4})). 
This expression provides quite different behavior of $\overline{k}(s,t)$ than that one that we considered in 
the previous section. 

\begin{figure}
\epsfxsize=85mm
\epsffile{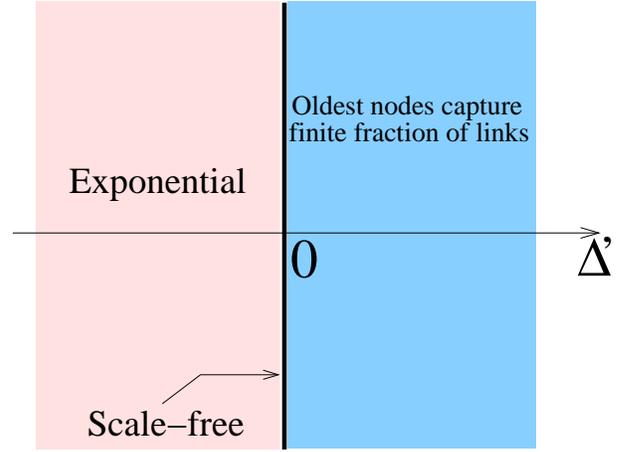}
\caption{
Phases of the network with the varying strength of nodes, $G(s) = s^{-\Delta^\prime}$. 
Scale-free networks are realized only at $\Delta^\prime=0$.
}
\label{f5}
\end{figure}

If $\Delta^\prime<0$, i.e., new nodes are ``stronger'' than 
old ones, we get $\kappa(s/t \to 0) \to 0$, and network is {\em exponential}.  
If $\Delta^\prime=0$, we get an ordinary scale-free network. 
If $\Delta^\prime>0$, then $\kappa(s/t)$ is extremely divergent function at $\xi=0$. 
This peculiarity is not integrated. It indicates that several oldest nodes capture a finite part of nodes. Our theory is not applicable in this case directly, since one can not solve a self-consistency 
equation and find $\beta$. We do not consider this situation in detail here. Such behavior may be compared with that one found for $G = G(k) = k^{y-1}$, $y>1$, where most of links turn to be attached to the oldest node \cite{krl00,kr00}. 
The ``phase diagram'' of the model is shown in Fig. \ref{f5}.

%%%%%%
%%%%%%
%%%%%%

\subsection{Local strong nodes}\label{ss-local}

Fluctuating strength of nodes, $G(s)$ is introduced in the resent papers of Bianconi 
and Barab\'asi, Ref. \cite{bb00a,bb00b}. 
%They called this characteristic ``fitness''. 
Using analogies with other cooperative problems of statistical mechanics, they indicate a new phenomenon arising in this situation. We, on the contrary, 
demonstrate the essence of evolution of such systems using the most simple (although quite general) example, without implementing any analogy. 

Let us start from the model with directed links introduced in Sec. \ref{ss-linear1}. To simplify the formulas, we assume that $A=0$ (one may see that this does not reduce extend of generality of the model which produces scaling exponents in the wide ranges of values, $2<\gamma<\infty$ and $0<\beta<1$). Let us recall the model (see Fig. \ref{f1},\,a).   
Each increment of time, a new node with $n$ attached to it links is added to the network. These $n$ links are directed to the new node. Simultaneously, $m$ extra links are distributed preferentially among nodes. The rule of preference is the same as in Sec. \ref{ss-linear1}, 
i.e., probability that a link is directed to a node $s$ is proportional to its in-degree, $q_s$ but with one exception -- one node, $\tilde{s}$, is ``stronger'' than others, that is  
probability that this node capture a preferentially distributed link is higher. It has an additional weight factor, $g>1$, and proportional to $g q_{\tilde{s}}$. This means that 
$G_s = 1 + (g-1)\delta_{s,\tilde{s}}$. 

Let us study behavior of the network at $t \gg \tilde{s}$.  
The equations for average in-degree are 

\begin{eqnarray}
& & 
\frac{\partial \overline{q}_{\tilde{s}}(t)}{\partial t} = m \frac{g \overline{q}_{\tilde{s}}(t)}{(g-1) \overline{q}_{\tilde{s}}(t) +\int_0^t ds\,\overline{q}(s,t)} \, , \ \overline{q}_{\tilde{s}}(t=\tilde{s}) = q_i\, ,
\nonumber 
\\[5pt] 
& & 
\frac{\partial \overline{q}(s,t)}{\partial t} = 
m \frac{\overline{q}(s,t)}{(g-1) \overline{q}_{\tilde{s}}(t) +\int_0^t ds\,\overline{q}(s,t)}\, , \
\overline{q}(t,t) = n 
\, .
\label{19--1}
\end{eqnarray}
Here, 
$\overline{q}_{\tilde{s}}(t) +\int_0^t ds\,\overline{q}(s,t) = (m+n)t + {\cal O}(1)$. One can see that this equality alone is not sufficient to determine the denumerators in Eq. (\ref{19--1}).  

Two different situations are possible. In the first one, $q_{\tilde{s}}(t)$ grows slower than $t$, and, at long times, the denominators are equal $(m+n)t$. Then, second line of Eq. (\ref{19--1}) is similar to Eq. (\ref{5--4}), and we get again the exponents, $\beta = m/(m+n) \equiv \beta_0$ and 
$\gamma = 2 + n/m \equiv \gamma_0 =1 + 1/\beta_0$, were $0<\beta_0<1, \ 2<\gamma_0<\infty$. (It is convenient to write equations using not the parameters, $m$ and $n$, but $\gamma_0$ or $\beta_0$, that is the exponents of the network in which all nodes have equal strength, $g=1$.) 
The first line of Eq. (\ref{19--1}), in this case, looks as 
 
\begin{equation}
\frac{\partial \overline{q}_{\tilde{s}}(t)}{\partial t} = 
\frac{g m}{m+n} \frac{g \overline{q}_{\tilde{s}}(t)}{t}
\, .
\label{19--2}
\end{equation}
Hence, at long times, 
$\overline{q}_{\tilde{s}}(t) = const(q_i)\, t^{gm/(m+n)}$, and we see that the in-degree of the strong node 
does grow slower than $t$ only for
 
\begin{equation}
g < g_c \equiv 1 + \frac{n}{m} = \gamma_0 - 1 = \beta_0^{-1} > 1 
\, ,
\label{19--3}
\end{equation} 
so we obtain the natural threshold value.

In the other situation, $g>g_c$, at long times, we have the only possibility, 
$q_{\tilde{s}}(t) = d\, t$, $d$ is some constant, since more rapid growth of $q_{\tilde{s}}(t)$ is impossible. $d < m+n$. This means that, for $g>g_c$, finite part of all preferentially distributed links is 
captured by the strong node (in \cite{bb00b} this situation is called condensation). We see that a single strong node may produce a macroscopic effect, so $g_c$ is a {\em giant capture threshold}.
In this case, Eq. (\ref{19--1}) takes the form, 

\begin{eqnarray}
& & 
\frac{\partial \overline{q}_{\tilde{s}}(t)}{\partial t} = 
\frac{gm}{(g-1)d + m + n}
\frac{\overline{q}_{\tilde{s}}(t)}{t}
\, ,
\nonumber 
\\[5pt] 
& & 
\frac{\partial \overline{q}(s,t)}{\partial t} = 
 \frac{m}{(g-1)d + m + n}
\frac{\overline{q}(s,t)}{t} 
\, .
\label{19--4}
\end{eqnarray}  
Note that the coefficient in the first line is always larger than the one in the second line, 
since $g_c>1$.  
From the first line of Eq. (\ref{19--4}), we get the condition 
 
\begin{equation}
\frac{gm}{(g-1)d + m + n} = 1 
\, ,
\label{19--4a}
\end{equation} 
so, for $g>g_c$, the following part of all links of the network is captured by the strongest node: 
 
\begin{equation}
\frac{d}{m+n} = \frac{d}{m}\frac{m}{m+n} = \frac{1}{g_c}\frac{g-g_c}{g-1}
\, .
\label{19--5}
\end{equation} 
We have to emphasize that $d$ is independent on initial conditions! (Recall that we consider the long time limit.) This {\em giant capture} leads to change of exponents. 
Using the condition Eq. (\ref{19--4a}), we get immediately the following expressions for them, 
 
\begin{equation}
\beta = \frac{1}{g} < \beta_0 \, , \ \ \ \gamma = 1 + g > \gamma_0
\, .
\label{19--6}
\end{equation} 

\begin{figure}
\epsfxsize=65mm
\epsffile{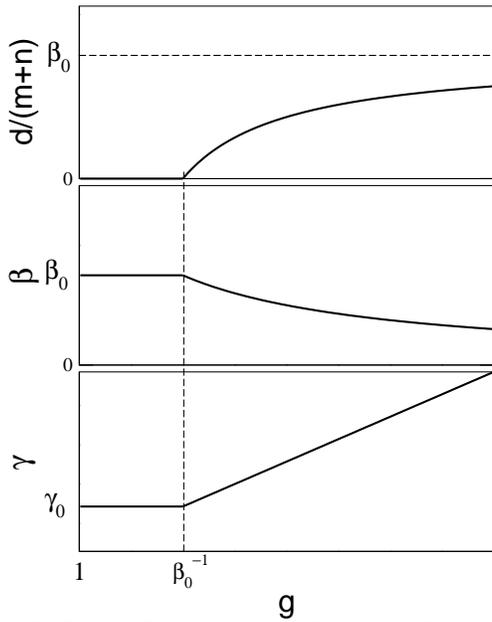}
\caption{ 
Giant capture. 
Fraction of all links, $d/(m+n)$, captured by a single strong node at long times and the scaling 
exponents $\beta$ and $\gamma$ vs relative strength, $g$, of the strong node. 
The giant capture occures above the threshold value, $g_c = 1/\beta_0 = \gamma_0-1>1$. 
Here, $\beta_0$ and $\gamma_0$ are the corresponding exponents for this network without the strong node. $d/(m+n)[g \to \infty] \to \beta_0$ and $\beta[g \to \infty] \to 0$. 
}
\label{f6}
\end{figure}

The fraction of all links captured by the strongest node and the $\beta$ and $\gamma$ exponents vs $g$ are shown in Fig. \ref{f6}.  
Note that growth of $g$ increases the value of the $\gamma$ exponent. If World is captured by Bill Geits or tzar, the distribution of wealth becomes more fair! 
One should note that the strong node does not take links away from other nodes but only {\em intercepts} them.
The closer $\gamma_0$ to $2$, the smaller $g$ is necessary to exceed the threshold. 
Above the threshold, values of the exponents are determined only by $g$. Nevertheless, the expression for $d/(m+n)$ contains $\gamma_0$ or $\beta_0$.

%%In \cite{xxx}, a similar phenomenon were studied in situation where at the threshold 
%%point $gamma=2$, and capture of the finite part of links is captured when $\gamma<2$. 
%%Such distribution itself leads to capture of a finite fraction of links by several nodes. 
%%%%This means that finite number of nodes capture a finite part of the 
%%%%distributed links, and it is hard to define number of ``condenced'' links. 
%%In our model, we have the capture accompaniented by increase of $\gamma$, 
%%and the process of the capture is defined quite clearly.   

\begin{figure}
\epsfxsize=85mm
\epsffile{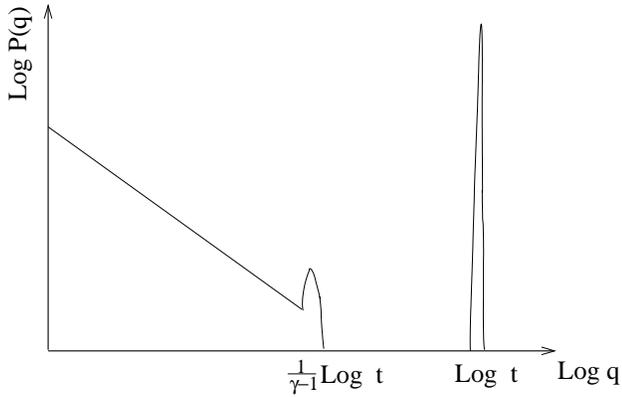}
\caption{
Schematic plot of the degree distribution of the network with one node which strength 
exceeds the threshold value. The giant peak is due to the strong node. A hump at the cut-off of the continuous part of the distribution is a trace of initial conditions (see Sec. 
\protect\ref{ss-cut-off} and Ref. \protect\cite{xxx}).
}
\label{f7}
\end{figure}

For $g>g_c$, in the giant capture regime, 
%%the strength of 
the strongest node
%%, $g$, 
determines the evolution of the network. With increasing time, a gap between the in-degree of the strongest node and maximal in-degree of all others grows, see Fig. \ref{f7}. A small peak at the end of the continuous part of the distribution is a trace of initial conditions, see Sec. \ref{ss-cut-off}. Note that the network remains scale-free even above the threshold, i.e., for $g>g_c$, although $\gamma$ grows with growing $g$.

The same result may be obtained for several nodes, $\tilde{s}_i$ with different strengths $g_i$, where
$\tilde{s}_i \ll t$. Here, $g_i > 1$ ($g_i<1$ do not produce any visible effect). In this case, the strongest node, $\tilde{s}_j$, again captures a finite part of nodes if $g_j>g_c$. 
Note that the time which the strongest node needs to seize a finite fraction of links may be very long if it is only a bit stronger than the previous strongest one. Note also that a death of the strongest guy produces a dramatic effect -- after some time, the distribution becomes less ``fair''. 
Deleting of a single strong node may also destroy entire network.

If $N$ strong nodes of equal strength, $g$, are present, one can find that, again, the critical exponents are described by Eq. (\ref{19--6}) and the part of links captured by all these strong nodes together is given by by Eq. (\ref{19--5}).

%%%%%%
%%%%%%
%%%%%%

\subsection{Mixture of ``weak'' and ``strong'' nodes}\label{ss-mix}

How one may ``smear'' the giant capture threshold described in the previous section? Again, let us use the simplest model in which nodes have two values of strength, $1$ and $g>1$, but now the nodes with different strengths are distributed homogeneously. Probability that a node has the strength equal $1$ is $1-p$, and, with probability $p$, a node has the strength $g$. All other conditions are the same. 
We can introduce average in-degrees, $\overline{q}_1(t)$ and $\overline{q}_g(t)$ for these components.  
Then, the evolution of the average in-degrees is determined by the following equations,  

\begin{eqnarray}
& & 
\frac{\partial \overline{q}_g(s,t)}{\partial t} = 
m \frac{g \overline{q}_g(s,t)}{(1-p)\int_0^t ds\,\overline{q}_1(s,t) + g\,p\int_0^t ds\,\overline{q}_g(s,t)}
\, , 
\nonumber 
\\[5pt] 
& & 
\frac{\partial \overline{q}_1(s,t)}{\partial t} = 
m \frac{\overline{q}_1(s,t)}{(1-p)\int_0^t ds\,\overline{q}_1(s,t) + g\,p\int_0^t ds\,\overline{q}_g(s,t)}
\, 
\label{20--1}
\end{eqnarray}
(compare with Eq. (\ref{19--1})). As usual, we get 

\begin{equation}
(1-p)\int_0^t ds\,\overline{q}_1(s,t) + p\int_0^t ds\,\overline{q}_g(s,t) = (m+n)t 
\, .
\label{20--2}
\end{equation} 
If one introduce the natural notation, 
$m/[(1-p)\int_0^t ds\,\overline{q}_1(s,t) + g\,p\int_0^t ds\,\overline{q}_g(s,t)] \equiv \beta_1$, 
Eq. (\ref{20--1}) takes the following form, 

\begin{eqnarray}
& & 
\frac{\partial \overline{q}_g(s,t)}{\partial t} = 
\beta_1 g \frac{\overline{q}_g(s,t)}{t}
\, , 
\nonumber 
\\[5pt] 
& & 
\frac{\partial \overline{q}_1(s,t)}{\partial t} = 
\beta_1 \frac{\overline{q}_1(s,t)}{t}
\, .
\label{20--3}
\end{eqnarray}
Inserting the solutions of these equations, $\overline{q}_g(s,t) = n(s/t)^{-\beta_1 g}$ 
and $\overline{q}_1(s,t) = n(s/t)^{-\beta_1}$ into 
%the link conservation relation, 
Eq. (\ref{20--2}), we obtain the equation for $\beta_1$, 

\begin{equation}
\frac{1-p}{1-\beta_1} + \frac{p}{1-\beta_1 g} = \frac{1}{1-\beta_0}
\, .
\label{20--4}
\end{equation} 
The solution of Eq. (\ref{20--4}) is of the form,

\begin{eqnarray}
& & 
\beta_1 = \frac{1}{2g} 
\{
[1-p+p\beta_0]+[p+(1-p)\beta_0]g - 
\nonumber 
\\[5pt]
& & 
\sqrt{\{ [1-p+p\beta_0]+[p+(1-p)\beta_0]g \}^2 - 4 \beta_0 g}\,
\}
\, 
\label{20--5}
\end{eqnarray} 
(we chose this root of the square equation to obtain the proper equality, $\beta_1(g=1)=\beta_0$).

\begin{figure}
\epsfxsize=85mm
\epsffile{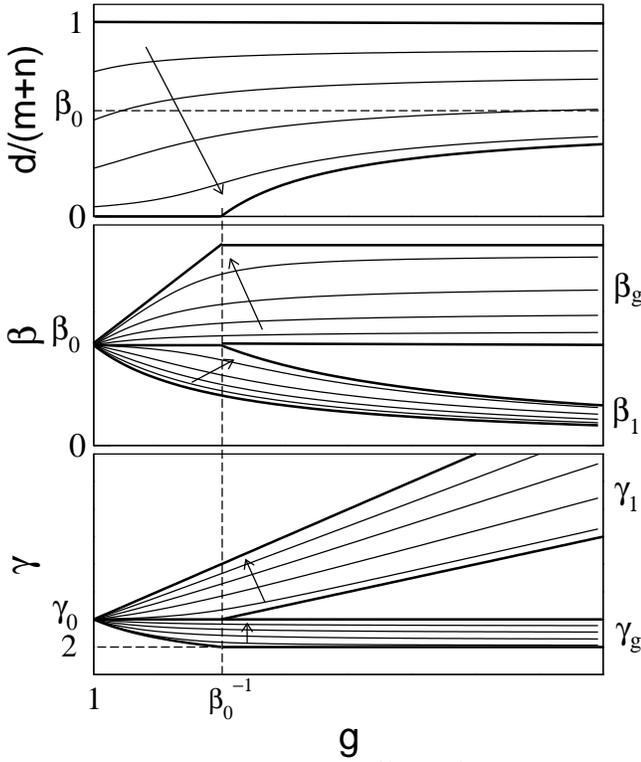}
\caption{
Fraction of all links, $d/(m+n)$, captured by the component of the network consisting of strong nodes,  at long times and the scaling 
exponents $\beta$ and $\gamma$ vs relative strength, $g$, of strong nodes. 
We introduce two sets of exponents for two components of the network, $\beta_1$ and $\gamma_1$ -- for the component consisting of nodes with the unit strength (contains $(1-p)t$ nodes) and $\beta_g$ and $\gamma_g$ -- for the component consisting of nodes with the strength $g$ (contains $pt$ nodes). 
Thin lines depict the dependences at fixed  values of $p$. 
Arrows show how these curves change when $p$ decreases from $1$ to $0$. At $p \to 0$, we obtain dependences shown in Fig. \protect\ref{f8}. 
At $p \to 1$, $d/(m+n) \to 1$, $\beta_g(g<g_c)\to \beta_0 g,\, \beta_g(g>g_c)\to 1$, 
$\beta_1 \to \beta_0/g$, $\gamma_1 \to 1+g/\beta_0$, $\gamma_g(g<g_c) \to 1+1/(\beta_0g),\, 
\gamma_g(g>g_c) \to \gamma_0$. 
}
\label{f8}
\end{figure}

The fraction of links captured by the strong component is given by the relation, 

\begin{equation}
\frac{d}{m+n} = \frac{p}{1-\beta_1 g}\,\frac{n}{m+n} =  p\frac{1-\beta_0}{1-\beta_1 g}
\, 
\label{20--6}
\end{equation} 
with $\beta_1$ taken from Eq. (\ref{20--5}). Exponents $\beta_1$ and $\gamma_1=1+1/\beta_1$ describes the distribution of nodes of the weak component containing the $1-d/(m+n)$ part of nodes. 
Exponents $\beta_g=\beta_1 g$ and $\gamma_g=1+1/\beta_g$ describes the distribution of nodes of the strong component. The dependences of these characteristics on $g$ are shown schematically in Fig. \ref{f8}. 
One may see that, for $p \to 0$ these curves tend to the dependences obtained for a single strong node in the previous section. For $p>0$, a long tail of the distribution with the exponent 
$\gamma_g < \gamma_0 < \gamma_1$ is determined by strong nodes. At $p \to 0$, it is transformed 
into the $\delta$-function obtained in the previous section. A particular value of $p$ determines the smearing of the giant capture threshold.

%%%%%%%%%%
%%%%%%%%%%
%%%%%%%%%%
%%%%%%%%%%

\section{Distributions of links between pairs of nodes}\label{s-density}

Let us discuss another characteristic describing a structure of networks -- distribution of links between pairs of nodes -- an average matrix element of an adjacency matrix. (An element $b_{i,j}$ of an adjacency matrix equals $1$ if there is a link connecting the sites $i$ and $j$ and it equals $0$ is the link is absent.) 

There is a very important particular case, when this characteristic may be may be easily calculated. In this case, new links appear only between new node and old ones but never -- between old nodes of a network (see Fig. \ref{f1},\,b). For instance, networks of scientific citations belongs to class of such networks.   

Here, we study the simplest model of Sec. \ref{ss-linear1} with $n=0$.
For brevity, we consider in this section networks with undirected links, so, into the relations of 
Sec. \ref{ss-linear1} we may substitute $\overline{q}(s,t) = \overline{k}(s,t) - m$. 
That is, at each time step, we add a new node with $m$ attached undirected links. Other ends of these links are distributed preferentially between old nodes.  

Let us introduce the average number of links between nodes $s$ and $s^\prime$ at time $t$, 
$\overline{b}(s,s^\prime,t)$. For a network that we consider,
 
\begin{eqnarray}
\overline{k}(s,t) & = & 
\int_0^s du\, \overline{b}(u,s,t) + \int_s^t du\, \overline{b}(s,u,t) = 
\nonumber
\\ [5pt]
& & m + \int_s^t du\, \overline{b}(s,u,t)
\, .
\label{10--1}
\end{eqnarray}
No new links between old nodes arise, so one can see that 
 
\begin{equation}
\overline{b}(s,t,t^\prime \geq t) = \overline{b}(s,t,t) \, , \ 
\frac{\partial \overline{k}(s,t)}{\partial t} = \overline{b}(s,t,t^\prime \geq t)
\, .
\label{10--2}
\end{equation}

Using scaling representation, $\overline{k}(s,t) = \kappa(s/t)$ and 
$\overline{b}(s,s^\prime,t) = t^{-1}{\cal B}(s/t,s^\prime/t)$, one can write 
 
\begin{equation}
\overline{b}(s,s^\prime,t^\prime \geq t) = 
\frac{1}{t^\prime} {\cal B}\left(\frac{s}{t^\prime},\frac{t}{t^\prime}\right) =
\frac{1}{t} {\cal B}\left(\frac{s}{t},\frac{t}{t}=1\right)  
%\, 
\label{10--3}
\end{equation}
and
 
\begin{equation}
\frac{\partial }{\partial (s/t)}\, \kappa\left( \frac{s}{t} \right) = 
\frac{\partial }{\partial (s/t)}\, \kappa\left( \frac{s}{t^\prime}\,\frac{t^\prime}{t} \right)
\, 
\label{10--4}
\end{equation} 
Therefore, we get the following relations for ${\cal B}(\xi,\xi^\prime)$ and $\kappa(\xi)$,
 
\begin{eqnarray}
{\cal B}\left(\xi,\xi^\prime\right) & = & 
\frac{1}{\xi^\prime} {\cal B}\left( \frac{\xi}{\xi^\prime}, 1 \right)
\, ,
\nonumber
\\[5pt]
{\cal B}\left(\xi,\xi^\prime\right) & = & 
- \frac{\xi}{\xi^{\prime\,2}}\,\kappa^\prime\left(\frac{\xi}{\xi^\prime}\right) 
\, ,
\nonumber
\\[7pt]
{\cal B}(\xi,1) & = & - \xi\, \kappa^\prime(\xi)
\, 
\label{10--5}
\end{eqnarray} 
where $\kappa^\prime(x) \equiv d\kappa(x)/dx$. 

In our particular problem, $\kappa(1)=m$, so, from Eq. (\ref{10--1}), it follows that
 
\begin{equation}
\kappa(\xi) = 
%\int_0^\xi d\xi^{\prime\prime}\, {\cal B}(\xi^{\prime\prime},\xi^\prime) +
%\int_\xi^1 d\xi^{\prime\prime}\, {\cal B}(\xi,\xi^{\prime\prime}) = 
m + \int_\xi^1 d\xi^{\prime\prime}\, {\cal B}(\xi,\xi^{\prime\prime}) = 
m + \int_\xi^1 \frac{d\zeta}{\zeta}\, {\cal B}(\zeta,1)
\, .
\label{10--6}
\end{equation} 

From relations of Sec. \ref{ss-linear1}, we obtain
 
\begin{equation}
\kappa(\xi) = 
m \left[2-\frac{1}{\beta} + \left( \frac{1}{\beta} - 1 \right)\xi^{-\beta} \right]
\, ,
\label{10--7}
\end{equation} 
so we get the result,  

\begin{equation}
{\cal B}(\xi,\xi^{\prime}) = 
m (1 - \beta) \xi^{-\beta} \xi^{\prime\, \beta-1}
\, .
\label{10--8}
\end{equation} 
This characteristic was obtained explicitly for a model with $\beta=1/2$ \cite{dms00c}. From this exact solution, the particular case, $\beta=1/2$, of Eq. (\ref{10--8}) follows immediately. Note that 
${\cal B}(\xi,\xi^{\prime})$ does not proportional to the product, $\kappa(\xi)\kappa(\xi^{\prime})$.

It is interesting to compare these expressions with the corresponding results for exponential networks (we put off the discussion of applicability of the continuous approach to non-scale-free networks until Sec. \ref{ss-random}). 
Setting $m=n_r$ and $n=A=0$ in the equations of Sec. \ref{ss-mixture} one gets

\begin{equation}
\kappa(\xi) = m(1 - \ln \xi) 
\, , \ \  \
{\cal B}(\xi,\xi^{\prime}) = 
\frac{m}{ \xi^{\prime}}
\, .
\label{10--9}
\end{equation}

Eq. (\ref{10--9}) follows from Eqs. (\ref{10--7}) and (\ref{10--8}) at $\beta \to 0$. 

We have obtained the density of linkage between pairs of nodes with fixed times of birth. 
From this characteristic (in the continuous approximation), we can find average number of connections, ${\cal D}(k,k^\prime)$, of parent nodes with degree, 
$k$, and child nodes with degree, $k^\prime$. In the continuous approximation, 
$k>k^\prime$. 

Let us consider only the scaling region and will not write coefficients in the following formulas of the present section. $k=\kappa(\xi)$, so, for a scale-free network,  
$P(k) = -(\partial\kappa/\partial \xi)^{-1} \sim k^{-1/\beta-1}$. Analogously, 

\begin{eqnarray}
& & 
{\cal D}(k,k^\prime) = 
\nonumber 
\\[5pt]
& & 
B(\xi,\xi^\prime) 
\left(\frac{\partial k}{\partial\xi}\right)^{-1}  \!
\left(\frac{\partial k^{\prime}}{\partial\xi^{\prime}}\right)^{-1} \!\!
[\xi=k^{-1/\beta},\xi^{\prime}=k^{\prime\,-1/\beta}] \propto 
\nonumber 
\\[5pt]
& & \phantom{xxxxxxxii}
k^{-1/\beta}k^{\prime\,-2}
\, .
\label{10--10}
\end{eqnarray} 
One sees that 
${\cal D}(k,k^\prime)$ is decreasing function of both $k$ and $k^\prime$. Therefore, 
if $k$ (parent node) is fixed, the most probable linking is with a child node with 
the smallest possible $k^\prime$, i.e., with the node with $k^\prime\sim 1$. 

 If $k^\prime$ (child node) is fixed, the most probable linking is with a parent node with 
the smallest possible $k$, i.e., with the node with $k=k^\prime$. 
This answer may be compared with the corresponding exact result of Krapivsky and Redner \cite{kr00} -- the maximum of the probability that a node with degree $k$ (parent) and a node with degree $k^\prime$ (child) are connected, when $k^\prime$ is fixed, is at $k/k^\prime=0.372$. 
Therefore, we see, that the continuous approach is good also in this situation. 

Eq. (\ref{10--10}) looks asymmetrically and cannot be factorized. The reason is the obvious absence of the symmetry relatively reversal of time -- 
a quite natural asymmetry between parents and childs. 

%%Let us repeat the calculations for the network growing with random linking 
%%(we delay discussion of the possibility of application of the continuous 
%%approach for such networks until Sec. \ref{ss-random}). 
%%
%%For such a network, using Eq. (\ref{10--9}), we obtain 
%%$P(k) = -(\partial\kappa/\partial \xi)^{-1} = m\exp(1-k/m)$ and 
%%${\cal P}(k,k^\prime) = \frac{1}{m} \exp(1-k/m)$. 
%%
%%It is better to remove the very last peace -- For exponential networks, 
%%${\cal P}(k,k^\prime)$, do not depend on $k^\prime$. For them, it may mean 
%%exponential dependence, and there is a chance that this simmetry may be restored for them.

%%%%%%%%%%
%%%%%%%%%%
%%%%%%%%%%
%%%%%%%%%%

\section{Accelerating growth}\label{s-accelerating}

Above we studied only linear network growth, i.e., the total number of links of a network divided by the total number of its nodes was constant during the evolution. This is only a particular case of network growth. The total number of links may be a nonlinear function of the total number of nodes. Keeping in mind the most intriguing applications, communications networks, we concentrate now on accelerating growth \cite{dm00c} where this function grows more rapidly than a linear one. In particular, we will see that a power-law dependence of input flow of links produces scale-free networks.

\subsection{Scaling relations}\label{ss-scaling2}

In the present section, we consider scale-free networks with power-law dependent on $t$ input flow of links. 
We will see that, in a large network size limit, such nonlinear growth may produce the following non-stationary distributions, 

\begin{equation}
P(q,t) \propto t^z q^{-\gamma}
\, ,  
\label{12--1}
\end{equation} 
and average indegree (or degree),

\begin{equation}
\overline{q}(s,t) \propto t^\delta \left( \frac{s}{t} \right)^{-\beta}
\, .   
\label{12--2}
\end{equation} 
Let us discuss what is the result of such non-stationary behavior.

One can easily repeat the derivation of Sec. \ref{s-scaling1} and obtain the scaling form of the connectivity distribution of individual nodes:

\begin{equation}
p(q,s,t) = t^{-\delta} \left( \frac{s}{t} \right)^\beta f\left[ q\, t^{-\delta} \left( \frac{s}{t} \right)^\beta \right]
\, .   
\label{12--3}
\end{equation} 
We again use the definition of $P(k,t)$, Eq. (\ref{2--7}), and Eq. (\ref{12--3}):

\begin{equation}
\int_0^\infty dx\, t^{-\delta} x^\beta f(q\,t^{-\delta}x^\beta) 
\propto t^{\delta/\beta}q^{-1-1/\beta} \propto t^z q^{-\gamma}
\, .   
\label{12--4}
\end{equation}
Hence, we get the universal scaling relation for the exponents $z$, $\delta$, and $\beta$, 

\begin{equation}
z = \delta/\beta
\,    
\label{12--5}
\end{equation}
and the old relation, (\ref{3--3}), for the exponents $\gamma$ and $\beta$. 

Let us derive this result once again applying the continuous approach. Using Eqs. (\ref{3--2}), 
(\ref{12--1}), and (\ref{12--2}), we obtain

\begin{equation}
t^z q^{-\gamma} \propto - \frac{1}{t}\frac{\partial s}{\partial t}[q=t^{\delta+\beta}s^{-\beta}] 
\propto t^{\delta/\beta} q^{-1-1/\beta}
\, .   
\label{12--6}
\end{equation} 
From Eq. (\ref{12--6}), the relations, Eqs. (\ref{3--3}) and (\ref{12--5}) follows immediately. 

One can write the relation, Eq. (\ref{12--3}) using only the $\gamma$ exponent: 

\begin{equation}
p(q,s,t) = \frac{s^{1/(\gamma-1)}}{t^{(1+z)/(\gamma-1)}} 
f\left( q\,\frac{s^{1/(\gamma-1)}}{t^{(1+z)/(\gamma-1)}}  \right)
\, .   
\label{12--7}
\end{equation} 
Eq. (\ref{12--7}) is direct generalization of Eq. (\ref{3--5}) obtained for a linearly growing networks.

%%%%%
%%%%%%%%
%%%%%

\subsection{Cut-off of connectivity distributions}\label{ss-cut-off}

Eq. (\ref{12--1}) is valid only in the limit of large network size (long times). 
Let us discuss briefly arising finite-size effects.

Account for the relation Eq. (\ref{12--2}) of Sec. \ref{ss-scaling2} together with the rapid decrease of $p(q,s,t)$ 
at large $q$ produces a cut-off of power law distributions at the characteristic value, 

\begin{equation}
q_{cut} \sim t^{\beta+\delta} = t^{\beta(1+z)} = t^{(1+z)/(\gamma-1)}
\, .
\label{17--1}
\end{equation} 
Here, we used the fact that $\overline{q}(s,t)$ is largest for the oldest node and the scaling relations between the exponents, Eqs. (\ref{3--3}) and (\ref{12--5}). It was shown that a trace of initial conditions at $q \sim q_{cut}$ may be visible in degree- and in-degree distribution measured for any network sizes \cite{dms00c}. 
Such a cut-off (and a trace of initial conditions) inputs restrictions for observations of power-law distributions since there are few huge networks in Nature.

%%This means
%%
%%In the particular case, $z=0$, 
%%
%%% Note that  Eq. (\ref{17--1}) is valid for $\gamma \geq 2$.
%%
%%+ reference to our exact result 

Let us obtain a general form of $P(q,t)$ for scale-free networks in scaling regime. 
Using the known scaling form of $p(q,s,t)$, Eq. (\ref{12--7}), we can write,

\begin{eqnarray}
P(q,t) = \frac{1}{t} \int_1^t ds\,p(q,s,t) \sim
\nonumber
\\[5pt]
t^z q^{-\gamma} \int_{qt^{-(1+z)/(\gamma-1)}}^{qt^{-z/(\gamma-1)}} dw w^{\gamma-1}f(w)
\, .
\label{17--2}
\end{eqnarray} 
Passing to the scaling limit, $q \to \infty, \ t \to \infty$, and $qt^{-(1+z)/(\gamma-1)}$ is fixed, we can replace the upper limit of the integral in Eq. (\ref{17--2}) by $\infty$. 
Then, we get immediately the following scaling form,

\begin{equation}
P(q,t) =  t^z q^{-\gamma} F(qt^{-(1+z)\beta}) =
t^z q^{-\gamma} F(qt^{-(1+z)/(\gamma-1)})
\, ,
\label{17--3}
\end{equation} 
where $F(w)$ is a scaling function. In the case of linearly growing network, the exponent $z$ is zero, so 
$P(q,t) = q^{-\gamma} F(qt^{-1/\beta}) = q^{-\gamma} F(qt^{-1/(\gamma-1)})$. This relation was obtained for an exactly solvable model \cite{dms00c}.

%%%%%
%%%%%%%%
%%%%%

\subsection{Scaling exponents}\label{ss-scalingexp}

Using relations obtained in Secs. \ref{ss-scaling2} and \ref{ss-cut-off} one can obtain 
general results for a nonlinear, accelerating growth of networks.

We start from the most general considerations. 
In the scale-free networks, a wide range of the connectivity distribution function is of the power-law form, $P(q) \propto t^z q^{-\gamma}$. It will be clear from the following that, to keep a network in the class of free-scale nets, the flow of new links has to be a power function of the number of nodes of a network, i.e., be proportional to $t^\alpha$. 

First, let us assume that the exponent of the distribution is less than two. The reasonable range is $1<\gamma<2$. To produce the restricted average connectivity (that is proportional to $t^\alpha$), the distribution has to have a cut-off at large $q$, $q_c \sim t^{(1+z)/(\gamma-1)}$ (see Sec. \ref{ss-cut-off}). $P(q) \propto t^z q^{-\gamma}$ for $q_l \sim t^x \lesssim q \lesssim q_c$. 
%$z > 0$ since the network grows. 
The restriction from below is necessary to garantee convergence of the integral, 
$\int_0^\infty dq\,P(q,t)=1$. From this, we get immediately $x=z/(\gamma-1)$. (Of course, this relation is valid also for $\gamma>2$.) 

The average connectivity, $\overline{q}$ is of the order 
$t^{\alpha+1}/t$, then 
$t^\alpha \sim \int^{t^{(1+z)/(\gamma-1)}} dq\,q\,t^z q^{-\gamma} \sim 
t^{-1+(1+z)/(\gamma-1)}$ (the value of the integral is determined by its upper limit). 
Therefore, $(1+z)/(\gamma-1)=\alpha+1$, so the cut-off of the distribution is of the order of the total number of links in the network. This is the maximal number of the problem, hence, in fact, the cut-off is absent. The expression for the $\gamma$ exponent, 
$\gamma= 1+(1+z)/(1+\alpha)$, follows from the last relation. 

Note that $\alpha$ is ``external'' exponent which governs the growth process.  
We have demonstrated that it is enough to know $\alpha$ and only one exponent of 
%all the ones figurated in the problem, 
$\gamma, \beta, z, \delta, x$ for finding all the others. 

Note also that we have to set $z<\alpha$ to keep the exponent $\gamma$ below $2$ as it was  
assumed. Also, one sees that the lower boundary for $\gamma$, $1+1/(1+\alpha)$, is approached for the stationary distribution, $z=0$. In this case, the form of the distribution is completely fixed by the accelerating growth, the exponent $\gamma$ depends only on $\alpha$.  

The other possibility is $\gamma>2$. In this case, the integral for the average connectivity is 
determined by its lower limit, 
$t^\alpha \sim \int_{t^{z/(\gamma-1)}} dq\,q\,t^z q^{-\gamma} \sim t^{z-z(\gamma-2)/(\gamma-1)}$. Hence, $\gamma=1+z/\alpha$, and $z>\alpha$. (Of course, this relation is not valid for $\alpha=0$.) Thus, we have described the possible forms of the connectivity distribution. 

 Let us demonstrate how these distributions may arise in the nonlinearly growing networks with the preferential linking. We introduce the simplest generalizations of the model of Sec. \ref{ss-linear1} to the case of the increasing input flow of links, $c_0 t^\alpha$, $\alpha>0$.

First, let us consider the case of constant initial attractivity, $A=const$. The equation for 
$\overline{q}(s,t)$ is

\begin{equation}
\frac{\partial\overline{q}(s,t)}{\partial t} = c_0\, t^\alpha 
\frac{\overline{q}(s,t)+A}{\int_0^t du\, [\,\overline{q}(u,t)+A\,]}
\, ,  
\label{13--1}
\end{equation}   
$\overline{q}(0,0)=0, \ \overline{q}(t,t)=n$. 
One may check that $\int_0^t du\, \overline{q}(u,t) = nt + c_0 t^{\alpha+1}/(\alpha+1)$. 
Inserting this relation into Eq. (\ref{13--1}) and solving the resulting equation, one gets

\begin{equation}
\frac{\overline{q}(s,t)+A}{n+A}  = 
\left[ \frac{ 1 + (n+A)(1+\alpha)t^{-\alpha}/c_0 }
{ 1 + (n+A)(1+\alpha)s^{-\alpha}/c_0 } \right]^{\!\! 1+1/\alpha} \!\!\!
\left( \frac{s}{t} \right)^{-(\alpha+1)}
\! .  
\label{13--2}
\end{equation} 
In the interval $[(n+A)(1+\alpha)/c_0]^{1/\alpha} \ll s \ll t$,

\begin{equation}
\overline{q}(s,t)  = (n+A)\,\left( \frac{s}{t} \right)^{-(\alpha+1)}
\, .  
\label{13--3}
\end{equation} 
Thus, the exponent $\beta$, $\overline{q}(s,t) \propto s^{-\beta}$, equals $1+\alpha$ and is larger than $1$. 
The dependence $\overline{q}(s)$ becomes constant, 

\begin{equation}
\overline{q}(s,t)  = 
(n+A)^{-1/\alpha} \left( \frac{c_0}{1+\alpha} \right)^{1+1/\alpha} t^{\alpha+1} 
\, ,
\label{13--4}
\end{equation} 
at $s \ll [(n+A)(1+\alpha)/c_0]^{1/\alpha}$. One may compare the result, Eq. (\ref{13--4}), with the total number of links in the network, $N(t) \approx c_0 t^{\alpha+1}/(1+\alpha)$. 

From Eq. (\ref{13--4}), we see immediately that the exponent $\beta=1+\alpha$, so 
$\gamma=1+1/\alpha$, $\delta=z=0$. One may calculate the connectivity distribution using Eq. (\ref{3--2}). The resulting distribution, in the region 
$1 \ll q/(n+A) \ll \left\{\, c_0/[(n+A)(1+\alpha)]\,\right\}^{1+1/\alpha}t^{1+\alpha}$, 
is of the form,

\begin{equation}
P(q,t) = \frac{(n+A)^{1/(1+\alpha)}}{1+\alpha}\, q^{-[1+1/(1+\alpha)]}
\, .
\label{13--5}
\end{equation} 
Thus, we obtain the stationary connectivity distribution with the $\gamma$ exponent less than $2$ that belongs to one of the types described above. 

To demonstrate the other possibility, $\gamma>2$, we consider below the model with a different rule of the distribution of new links. Let the additional attractivity be time-dependent, 
and new links be distributed between nodes with probability proportional to 
$q + B c_0 t^{\alpha}/(1+\alpha)$, where $B$ is positive constant. $c_0 t^{\alpha}/(1+\alpha)$ is the average connectivity of the network at time $t$. 

Repeating the previous calculations, one gets the equation 

\begin{equation}
\frac{\partial\overline{q}(s,t)}{\partial t} = c_0\, t^\alpha 
\frac{\overline{q}(s,t) + B c_0\, t^{\alpha}/(1+\alpha)}
{nt + B c_0\, t^{\alpha+1}/(1+\alpha) + c_0 t^{\alpha+1}/(\alpha+1) } 
\label{13--6}
\end{equation}
$\overline{q}(0.0)=0,\ \overline{q}(t,t)=n$. At long times, one obtains

\begin{equation}
\frac{\partial\overline{q}(s,t)}{\partial t} = 
\frac{1+\alpha}{1+B}\, 
\frac{\overline{q}(s,t) + B c_0\, t^{\alpha}/(1+\alpha)}{ t } 
\, .
\label{13--7}
\end{equation}
The solution of Eq. (\ref{13--7}) is 

\begin{equation}
\overline{q}(s,t) = 
\left[ n + \frac{B c_0\, s^\alpha}{1-B\alpha}
%\,s^\alpha  
\right] 
\left( \frac{s}{t} \right)^{-(1+\alpha)/(1+B)}
- \frac{B c_0\,  t^\alpha}{1-B\alpha}
%\, t^\alpha
\, 
.
\label{13--8}
\end{equation}
If $B=0$, we obtain the previous result, $\beta=1+\alpha$. 
For $s^\alpha \gg n(1-B\alpha)/(B c_0)$, 

\begin{equation}
\overline{q}(s,t) \approx  
\frac{B c_0 t^\alpha}{1-B\alpha} 
\left\{ \left(\frac{s}{t} \right)^{\alpha-(1+\alpha)/(1+B)} - 1 \right\} 
\, .
\label{13--9}
\end{equation}
Therefore, the scaling exponents of the growing network are 
$\beta = (1+\alpha)/(1+B) - \alpha = (1-B\alpha)/(1+B)$,  
$\gamma = 1+1/\beta = 1 + [ (1+\alpha)/(1+B) - \alpha ]^{-1} = 
2 + B(1+\alpha)/(1-B\alpha)$, $\delta=\alpha$, and $z = \alpha(1+B)/(1-B\alpha)$. 
The connectivity distribution 
differs sharply from the distribution obtained for the previous model. 
It is nonstationary and 
is of the form $P(q,t) \sim t^{\alpha(1+B)/(1-B\alpha)}q^{-[1 + (1+B)/(1-B\alpha)]}$ for $q \gg t^\alpha$.
In this case, $\beta<1$ and $\gamma>2$ for any positive $\alpha$ and 
$B$. The scaling regime is realized if $B\alpha<1$. 

Note  
that, in both considered cases, one cannot set $\alpha=0$ directly in 
the obtained expression for the scaling exponents.

%%%%%%%%%%
%%%%%%%%%%
%%%%%%%%%%
%%%%%%%%%%

\section{Developing and decaying networks}\label{s-developing} 

Now we can study evolution of network accompanied by reconstruction of its old part \cite{dm00b}. This, e.g., may include permanent deletion of old links or nodes. Note that the processes of addition and deletion of links may be considered in a unified way, so we study them together. 
For brevity, in the present section, we consider undirected links.

%%%%%%
%%%%%%
%%%%%%

\subsection{Developing networks}\label{ss-developing}

Let us introduce two channels of appearance of new links. The first one was studied in 
Sec. \ref{ss-linear1}. We consider undirected links, so, in our old formulas, we have to 
substitute $\overline{q}(s,t) = \overline{k}(s,t) - m$ and put $\overline{k}(t,t)=m$. 
$n=0$. Instead of the additional attractiveness $A$ we use here the constant $A_n = A - m>-m$. The second channel is the following.  
Each time a new node is appeared, $c$ additional links arise between old unconnected nodes $i$ and $j$ with probability proportional to the product $(k_i+A_o)(k_j+A_o)$, $A_o>-m$. 
Note that $A_n$ and $A_o$ may be not equal, so, here, we have a mixture of different preferences like in Sec. \ref{ss-mixture}.   
Then the equation for $\overline{k}(s,t)$ is of the form,

\begin{equation}
\frac{\partial \overline{k}(s,t)}{\partial t} = 
m \frac{\overline{k}(s,t)+A_n}{\int_0^t du\,[\overline{k}(u,t)+A_n]} + 
2c \frac{\overline{k}(s,t)+A_o}{\int_0^t du\,[\overline{k}(u,t)+A_o]}
\, .
\label{11--1}
\end{equation}
(see \cite{dm00b}). 
$\int_0^t ds\,\overline{k}(s,t) = 2(m+c)t$, so we obtain immediately the scaling exponent $\beta$,

\begin{equation}
\beta = \frac{m}{2m+2c+A_n} + \frac{2c}{2m+2c+A_o}
\, 
\label{11--2}
\end{equation} 
and using the relation Eq. (\ref{3--3}), the exponent $\gamma$. 
%When $c$ is increased from $0$ to $\infty$, $\gamma$ decreases from the $\gamma(c=0)$ 
%value to $2$. 
We do not consider these relations in detail.

%%%%%%%%%%%%%
%%%%%%%%%%%%%
%%%%%%%%%%%%%

\subsection{Decaying networks}\label{ss-decaying}

Let us introduce possibility of deleting of links. 
The second channel now is the following. Each increment of time $-c$ random links are removed from the network. Note that we we save the same definition of the $c$ as in Sec. \ref{ss-developing} to use similar relations.

Previously, a process of instant random damage was considered 
\cite{ba00a,ceah00,cnsw00,seah00a}. 
Such type of damage can not change a value of $\gamma$ neither for random removal links nor for removal nodes.
Here, we consider a quite different situation, {\em permanent random damage} -- 
components of a network are removed permanently during its growth. In this case, difference between random removal of nodes and links is striking.  

We have shown that Eqs. (\ref{11--1}) and (\ref{11--2}) also describe the case of decaying networks 
with permanent deletion of links if one takes $c$ to be negative in them and assumes that $A_o=0$ \cite{dm00b}. 
%%(Compare with our exactly solvable model). 
Then, it follows from Eq. (\ref{11--2}) that 
 
\begin{eqnarray}
\beta = & & \frac{1}{2+2c/m+A_n/m} + \frac{c/m}{1+c/m} = 
\nonumber
\\[5pt]
& & 
\frac{1+\gamma_0 c/m+2(c/m)^2}{(1+c/m)(\gamma_0-1+2c/m)}
\, 
\label{11--3}
\end{eqnarray} 
and 
 
\begin{equation}
\gamma = 2 + \frac{1+\gamma_0 c/m+2(c/m)^2}{\gamma_0-2+c/m}
\, ,
\label{11--4}
\end{equation} 
where $\gamma_0 \equiv \gamma(c=0) = 3+A_n/m$. 

\begin{figure}
\epsfxsize=85mm
\epsffile{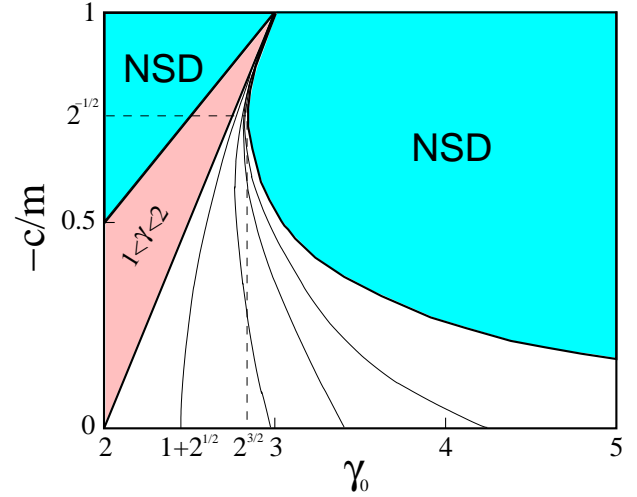}
\caption{
Phase diagram of the network growing in condition of permanent random damage -- 
permanent deleting of random links -- in axis $\gamma_0$, $-c/m$. At each time step, $m$ new links are added and 
$-c$ random links are deleted (see the text). 
$\gamma_0$ is the scaling exponent of the corresponding network growing without deleting of links. Curves in the plot are lines of constant values of $\gamma$. 
$\gamma=\infty$ on the line, $\gamma_0 = -2(c/m)-1/(c/m)$. 
When $1+\sqrt2<\gamma_0<2\sqrt2$, there is a maximum of the dependence, 
$\gamma(-c/m,\gamma_0 \mbox{fixed})$, at $-c/m=(1+\sqrt2)[\gamma_0-(1+\sqrt2)]/\sqrt2$. 
Between the lines, $-c/m=\gamma-2$ and $-c/m=(\gamma_0-1)/2$, the $\gamma$ exponent varies between the values $2$ and $1$ correspondingly. The NSD regions denotes areas where stationary distributions are absent.
}
\label{f9}
\end{figure}

The resulting phase diagram, $-c/m$ vs $\gamma_0$ is surprisingly rich, see Fig. \ref{f9}.  
One sees that the effect of deletion of links during the network growth may be quite different for different values of the $\gamma_0$ exponent of the corresponding network without such deletion. 
For $\gamma_0<1+\sqrt2$, the removal of links decreases $\gamma$. 
If $1+\sqrt2<\gamma_0<2\sqrt2$, deleting of links first increases $\gamma$ but afterwards $\gamma$ decreases. 
For $\gamma_0>2\sqrt2$, $\gamma$ grows monotonously with increasing $-c/m$ until it becomes infinite on the line, $\gamma_0 = -2(c/m) - 1/(c/m)$. We have checked our results by simulation in the particular case of $\gamma_0=3$ \cite{dm00b}. 

One should note that permanent deletion of links produces a region between the lines, $-c/m=(\gamma_0-1)/2$ and $-c/m=\gamma_0-2$, in which $\gamma$ varies between $1$ and $2$. It is the first time we find these distributions, with $1<\gamma<2$, in such problems.
In dashed regions of Fig. \ref{f9}, stationary distributions are absent. Note that, for large enough $-c/m$, the network may decay for a set of uncoupled clusters.

Let us consider now additional permanent deletion of random nodes. {\em Ab initio}, one may think that this factor does not change $\gamma$. Nevertheless, as we shall see,this case is of a special interest.  

At each increment of time, one new node is added, and a randomly chosen node is deleted with probability $c \leq 1$. We use the preferential linking introduced in Sec. \ref{ss-linear1}, 
although a more general model may be also considered. It is natural to consider in-degree, 
$q(s,t)$, here, since results are more general in this case.

Let us introduce the probability that a nodes $s$ is present, ${\cal N}(s,t)$. In the continuous approach it has meaning of ``density'' of surviving nodes at time $t$, 
$\overline{{\cal N}}(s,t)$. One may introduce ``density of in-degree''. In the continuous approach, it looks as 
 
\begin{equation}
\overline{d}(s,t) \equiv \overline{{\cal N}}(s,t) \overline{q}(s,t)
\, .
\label{11--9}
\end{equation}  

Our main equation for $\overline{{\cal N}}(s,t)$ is of the form,
 
\begin{equation}
\frac{\partial \overline{{\cal N}}(s,t)}{\partial t} = 
-c\, \frac{\overline{{\cal N}}(s,t)}{\int_0^t du\, \overline{{\cal N}}(u,t)}
\, ,
\label{11--10}
\end{equation}  
where $\overline{{\cal N}}(0,0)=0$ and $\overline{{\cal N}}(t,t)=1$. 
From Eq. (\ref{11--10}) one gets immediately the obvious relation, 
$\int_0^t du\, \overline{{\cal N}}(u,t) = (1-c)t$.
The solution of Eq. (\ref{11--10}) is 
 
\begin{equation}
\overline{{\cal N}}(s,t) = 
\left(\frac{s}{t}\right)^{c/(1-c)}
\, .
\label{11--11}
\end{equation}  

In the present case, the equation for the in-degree is 
 
\begin{equation}
\frac{\partial \overline{q}(s,t)}{\partial t} = 
m\, \frac{\overline{q}(s,t)+A}{\int_0^t du\, \overline{{\cal N}}(u,t)[\overline{q}(s,t)+A]}
\, ,
\label{11--12}
\end{equation}  
where the integral on the right side is equal to the sum of the total in-degree of the network at time $t$ and the product of the additional attractiveness and the number of sirvived nodes  (compare with Eq. (\ref{5--1}) for $c=0$). Here, $\overline{q}(0,0)=0$ and 
$\overline{q}(t,t)=n$ so $\overline{d}(0,0)=0$ and 
$\overline{d}(t,t)=n$. 

Multiplying Eq. (\ref{11--12}) by $\overline{{\cal N}}(s,t)$ and applying $\int_0^t ds$ to both sides of the resulting equation, we get 
$\int_0^t ds\, \overline{d}(s,t) = (m+n)(1-c)t$. Substituting this relation into Eq. (\ref{11--12}), we get immediately $\overline{q}(s,t) \propto (s/t)^{-\beta}$, where the exponent is
 
\begin{equation} 
\beta = \frac{m}{(1-c)(m+n+A)} = \frac{\beta_0}{1-c}
\, .
\label{11--13}
\end{equation}  
Here, $\beta_0 \equiv \beta(c=0), \ \gamma_0 \equiv \gamma(c=0)$. 
Using Eq. (\ref{11--11}), we obtain for the density of in-degree, 
$\overline{d}(s,t) \propto (s/t)^{-(\beta+c)/(1-c)}$. Note that, 
$\overline{d}(s \to 0,t) \to \infty$ for $c<\beta_0$, and 
$\overline{d}(s \to 0,t) \to 0$ for $c>\beta_0$. 

The expression for the degree distribution looks as 
 
\begin{eqnarray} 
P(k,t) & = & 
\frac{\int_0^t ds\, {\cal N}(s,t) p(q,s,t)}
{\int_0^t ds\, {\cal N}(s,t)} = 
\nonumber
\\[5pt]
& & 
\frac{\int_0^t ds\, \overline{{\cal N}}(s,t) \delta(q - \overline{q}(s,t))}
{\int_0^t ds\, \overline{{\cal N}}(s,t)}
\, .
\label{11--14}
\end{eqnarray} 
Therefore, repearting the derivation of Sec. \ref{s-scaling1} we obtain the following distribution,
$P(k) \propto k^{-(1/\beta)[c/(1-c)]} k^{-1-1/\beta} \propto k^{-1-1/[\beta(1-c)]}$, 
so $\gamma = 1 + 1/[\beta(1-c)] = 1 + 1/\beta_0 = \gamma_0$. 
Thus, the distribution is of the same form as without the permanent random deleting of nodes. 
Note that there is a great difference from the case of the permanent random deletion of links considered above. 

Here, we find for the first 
time a violation of the scaling relation, Eq. (\ref{3--3}). The reason is effective renormalization of the $s$ variable due to the removal of nodes. 

Repeating scaling considerations of previous sections, we get for this case the following forms of $p(q,s,t$ and $P(q)$, 
$p(q,s,t) = (s/t)^\beta f[ q (s/t)^\beta ]$ and $P(q) = q^{-\{1+1/[\beta(1-c)] \}} F(q/t^\beta)$.

We conclude this section by the statement that 
permanent deleting of a part of nodes with the largest values of degree, that is an analogy of intentional attacks \cite{ba00a,ceah00a}, destroys 
scaling behavior of a network. One may easily check this statement using the continuous approach.

%%%%%%%%%%
%%%%%%%%%%
%%%%%%%%%%
%%%%%%%%%%

\section{
%Area of 
applicability of the continuous approach}\label{s-applicability}

In the present section we discuss quality of the used continuous approach. For this, we compare known exact results with results obtained in frames of the continuous approximation.

%%%%%
%%%%%%%%
%%%%%

\subsection{Linear preference}\label{ss-linear2}

In Sec. \ref{s-reasons}, we have written out already the answer of the continuous approach for 
the Barab\'{a}si-Albert's model at $m=1$. For a more general case when $m$ is any positive integer number, and $k=q+m$, in the frames of the continuous approach, one gets, $P(q) = 2m^2 /(q+m)^3$ \cite{baj99}. 
One may compare this expression with the exact result obtained without passing to continuous limit: 
$P(q)=2m(m+1)/[(q+m)(q+m+1)(q+m+2)]$ \cite{dms00}. Exponents are the same but the factors are different. 

One may ask, why the approach is so good. 
The reason is rapid decrease of $p(q,s,t)$ at large $q$. 
Because this, the results of the continuous approximation obtained with the $\delta$-function ansatz, Eq. (\ref{2--5}), are reasonable.  
Indeed, in the limit of large $s$ and $t$ and for fixed $s/t$, we obtained the following 
scaling exponentially decreasing expression, $p(q,s,t) = [q(s/t)^\beta]^{A-1} \exp[-q(s/t)^\beta]/\Gamma(A)$ \cite{dms00}. 
The model of Sec. \ref{ss-linear1} with $n=0$ was considered. Here $\Gamma(\ )$ is the gamma-function.

%%%%%
%%%%%%%%
%%%%%

\subsection{Nonlinear preference}\label{ss-nonlinear}

The continuous approach may be applied to nonlinear preference. In this case, calculations similar to that ones of Sec. \ref{s-aging} may be made. 

Let us consider the simplest case, generalizing the Barab\'{a}si-Albert's model, 
$\overline{k}(t,t) = 1, m=1$. Then, the main equations are, 

\begin{equation}
\frac{\partial \overline{k}(s,t)}{\partial t} = 
\frac{f_p[\overline{k}(s,t)]}{\int_0^t du\, f_p[\overline{k}(u,t)]} \, \Longrightarrow \,
\int_0^t ds\,\overline{k}(s,t) = 2t
\, .  
\label{15--1}
\end{equation}
Here, $f_p(k)$ is a preference function.
Let us search for the solution of Eq. (\ref{15--1}) in the scaling form, $\overline{k}(s,t)=\kappa(s/t)$. Then,

\begin{equation}
-\frac{\partial \kappa(\xi)}{\partial \ln \xi} = 
\frac{f_p[\kappa(\xi)]}{\int_0^1 d\zeta\, f_p[\kappa(\zeta)]} \, , \ \ \ \kappa(1) = 1\, , \ \ \ 
\int_0^1 d\zeta\,\kappa(\zeta) = 2
\, .  
\label{15--2}
\end{equation}

Let us start from the case of nonlinear preference that produces scale-free networks. Let the probability for distribution of new links be proportional 
to the preference function, $f_p(k)$, that is linear asymptotically for large $k$, 
$f_p(k \to \infty) \to ck$, where $c$ is a constant. (see \cite{krl00,kr00}).

The integral $\int_0^1 d\zeta\, f_p[\kappa(\zeta)]$ is a constant of the problem. In the scaling region of large $\overline{k}(s,t)$ and $\kappa$, the equation takes the form, 

\begin{equation}
-d\ln\kappa(\xi)/d\ln\xi = c\left\{\int_0^1 d\zeta\, f_p[\kappa(\zeta)]\right\}^{-1} = \beta
\, . 
\label{15--2a}
\end{equation}
Eq. (\ref{15--2a}) demonstrates that the scaling is present, and the network is scale-free in this case. 

Therefore, to find the scaling exponent $\beta$, we have to solve the equation,

\begin{equation}
-\frac{\partial \kappa(\xi)}{\partial \ln \xi} = \frac{\beta}{c} f_p[\kappa(\xi)]
\,   
\label{15--3}
\end{equation} 
and after inserting its solution, 
$\kappa(\xi) = F^{-1}[F(1) + (\beta/c)\ln\xi]$ 
($F(\kappa) \equiv \int d\kappa\,f(\kappa)$, $F^{-1}$ is an inversed function to $F$), 
into 

\begin{equation}
\beta^{-1} = c^{-1}\int_0^1 d\zeta\, f_p[\kappa(\zeta)] \ \mbox{ or equivalently} \ 2 = \int_0^1 d\zeta\,\kappa(\zeta)
\,   
\label{15--4}
\end{equation}
find the solution $\beta$ of any of these transcendental equations. 

We will not consider examples of application of these relations but describe briefly the case of a power law preference function just to test the continuous approach using the known non-scale-free network. Indeed, we have checked the quality of the continuous approach for scale-free networks. Now it is naturally to check it for other networks. 
 
Let the preference function be $f_p(k) = k^{-y}$. 
If one sets

\begin{equation}
\int\limits_0^1 d\zeta \kappa^y(\zeta) =\mu = const 
 \, ,
\label{15--5}
\end{equation}
then the equation for $\kappa(\xi)$ is 

\begin{equation}
-\frac{d\ln \kappa(\xi)}{d\ln\xi}  = \mu^{-1}\exp[-(1-y)\ln\kappa]
 \, .
\label{15--6}
\end{equation}
Its solution is 

\begin{equation}
\kappa(\xi) = (1 -\frac{1-y}{\mu} \ln\xi)^{1/(1-y)}
\,  
%\ \ \ \ \ 
%\xi =
%\exp[\frac{\mu}{1-y}(1-\kappa^{1-y})]
 \, .
\label{15--7}
\end{equation} 
The constant $\mu$ can be obtained from the transcendental equations 

\begin{equation}
2 = \int\limits_0^1 d\xi \left(1 - \frac{1-y}{\mu}\ln\xi   \right)^{1/(1-y)}
\,
\label{15--8}
\end{equation}
or equivalently

\begin{equation}
\mu = \int\limits_0^1 d\xi \left(1 - \frac{1-y}{\mu}\ln\xi   \right)^{y/(1-y)}
\, .
\label{15--9}
\end{equation}

The final transcendental equation may be written in the form:

\begin{equation}
2\left(\frac{\mu}{1-y} \right)^{1/(1-y)}
 e^{-\mu/(1-y)} =
\Gamma\left(1+\frac{1}{1-y},\frac{\mu}{1-y}  \right)
\label{15--10}
\end{equation}
which gives $\mu(y)$. Near $y=1$, $\mu\cong2y$, near $y=0$, 
%%$\mu\cong 1+ 0.596347 y$, where $0.596347=e Ei(-1)$. 
$\mu\cong 1+ 0.5963 y$, where $0.5963 =e Ei(-1)$.

Inserting the solution, Eq. (\ref{15--7}), into the expression for $P(k)$ in the continuous approach, Eq. (\ref{3--2}) we obtain the connectivity distribution,

\begin{equation}
P(k) 
%\propto -\frac{d\xi(k)}{dk} 
= 
\mu e^{\mu/(1-y)} k^{-y} 
\exp\left[-\frac{\mu}{1-y} k^{1-y} \right] 
\, ,
\label{15--11}
\end{equation}
These results are close to the exact ones \cite{krl00,kr00}. The values of the powers are the same, although the coefficients, $\mu$ differs slightly. E.g., in Refs. \cite{krl00,kr00} 
Near $y=1$, $\mu\cong2 - 2.407(1-y)$, and, near $y=0$, 
$\mu\cong 1+ 0.5078 y$.

%%%%%
%%%%%%%%
%%%%%

\subsection{Random linking}\label{ss-random}

Finally, in the frames of the continuous approach, let us consider the model of Sec. \ref{ss-absence} producing non-scale-free, exponential degree distribution. 
This network is a continuous case of the network considered in Sec. \ref{ss-mixture}, $n=0, m=0, n_r=1, A=0$. The solution of Eq. (\ref{6--1}) in this particular case is

\begin{equation}
\overline{k}(s,t) = 1 - \ln(s/t)
\, 
\label{16--1}
\end{equation} 
(we used the boundary condition $\overline{k}(t,t) = 1$). 
Then, using Eq. (\ref{3--2}), one gets the connectivity distribution,

\begin{equation}
P(k) = - \frac{1}{t} \frac{\partial [t \exp(1-k)]}{\partial k} = \exp(1-k)
\, .
\label{16--2}
\end{equation} 
This can be compared with the exact form of Sec. \ref{ss-absence}, $P(k) = \exp(-k\ln 2)$.

Therefore, the continuous approach easily produces reasonable answers even for non-scale-free networks. The reason is again rapid decrease of $p(k,s,t)$ at large $k$ for these networks, see Eq. 
(\ref{4--4}) of Sec. \ref{ss-absence}.

%%----------
%%
%%Note, that if we get independent on $q$ distribution, in the continuous 
%%approach, it may mean that the distribution is exponential. 
%%
%%---------- 

%%%%%%%%%%
%%%%%%%%%%
%%%%%%%%%%
%%%%%%%%%%

\section*{CONCLUSIONS}

We have analysed scale-invariant properties of scale-free networks with growth governed by a mechanism of preferential linking. 
Degree distributions of such networks are of a simple scaling form. 
We have shown that the arising scaling exponents are coupled by universal scaling relations. Nevertheless, we present an important 
particular case in which these simple relations are violate. 

One of discussed questions has been about types of preferential linking producing scale-free networks. 
One can see that scale-free networks are produced by wide variety of linking. In particular, it is enough to 
add an admixture of linear preference linking to random linking to get a scale-free network. 
Interplay of different factors such as deletion of links of a network during its growth may change dramatically 
its degree distribution and even remove it from the class of scale-free networks. Hence, we have shown, how one 
can increase or decrease critical exponents of a network. 

All the time, we used the very simple continuous approach. Why it is so good? 
Scaling behavior of networks arises from a power law singularity of degree at $s=0$, i.e., for the oldest nodes, 
$\overline{k}(s,t) \propto s^{-\beta}$. Such, ``the oldest are the richest'', behavior is often perceived as a defect 
of a preferential linking scheme \cite{ah00}. If one removes such singularity or makes it more weak, a growing network 
will be out of the class of scale-free nets. It is the separation of power and exponential dependences in $P(q,t)$ and 
$p(q,s,t)$ that makes the continuous approximation so efficient. Therefore, we see that scale-free networks are quite 
suitable for the continuous approach. 

Natural boundaries, $s=0$ and $s=t$, are always present in growing network. We showed, that even presence of strong 
nodes did not lead to violation of the rule ``the oldest are the richest'' for nodes in the continuous part of 
degree-distribution. Nevertheless, we have found a threshold value of the strength of this node above that a 
single node influences evolution of an entire network. It captured a finite fraction of all links -- giant capture -- and 
determines values of exponents, although the network stays scale-free. These ``collective'' effects are explained by absence 
of any ``interaction distance'' in the process of network growth. Each node has some chance to get a new link. Therefore, 
in principle, it is possible to adjust parameters to direct all new links to a single node. 

One should note that main part of our paper has been devoted to studying of a one node characteristic of a network -- degree 
distribution. The same simple characteristic is a matter of interest of most of modern experimental and theoretical studies. 
This restriction let us to apply usually rather general models in which links appeared between arbitrary nodes: new and old, 
old and old. If we do not study statistics of connections between different nodes and global connectivity properties of a 
network, usually the considered problems are equivalent to a classical problem of distribution of new particles between 
increasing number of boxes.   
In fact, in the present paper, we studied the question, how a network is self-organized into a scale-free structure using 
some versions of stochastic multiplicative processes 
\cite{stoch} which are most known by their ``econophysical'' applications. Most of our results may be described both in 
terms of the theory of evolving networks and econophysics -- wealth distribution processes. 

Nevertheless, a part of our paper has been devoted to problems that cannot be understood using a one node characteristic. 
In this intriguing direction is, perhaps, the main aim of statistical physics of growing networks -- description of topology
 of evolving networks.

\cite{hppj98,ajb99,ha99,ls98,r98,ba99,ba00a,asbs00,mn00,b1,krrt99,fff99,bkm00}
\cite{baj99,dms00,krl00,dm00,dm00b}
\cite{ah00,note1,ceah00,er60,s55,be00}

.
\\

SND thanks PRAXIS XXI (Portugal) for a research grant PRAXIS XXI/BCC/16418/98. JFFM 
was partially supported by the project FCT-Sapiens 33141/99. 
We also thank A.N. Samukhin for many useful discussions. 
\\
$^{\ast}$      Electronic address: sdorogov@fc.up.pt\\
$^{\dagger}$   Electronic address: jfmendes@fc.up.pt

\newpage

%\newpage

\newpage

%\newpage

\end{multicols}

\end{document}